\documentclass[graybox]{svmult}

\usepackage{mathptmx}       
\usepackage{helvet}         
\usepackage{courier}        
\usepackage{type1cm}        
\usepackage{makeidx}         
\usepackage{graphicx}        
\usepackage{multicol}        
\usepackage[bottom]{footmisc}

\makeindex             

\begin{document}

\title*{Observational and Physical Classification of Supernovae}
\author{Avishay Gal-Yam}
\institute{Avishay Gal-Yam \at Department of Particle Physics and Astrophysics, Weizmann Institute of Science, Rehovot, Israel, \email{avishay.gal-yam@weizmann.ac.il}}
%
%
\maketitle


\abstract{This chapter describes the current classification scheme of supernovae (SNe). This scheme has evolved over many decades 
and now includes numerous SN Types and sub-types. Many of these are universally recognized, while there are controversies regarding
the definitions, membership and even the names of some sub-classes; we will try to review here the commonly-used nomenclature, noting 
the main variants when possible. SN Types are defined according to observational properties; mostly visible-light spectra near maximum 
light\footnote{All spectra presented and
discussed here are available from the public SN spectroscopy archive WISeREP (Yaron \& Gal-Yam 2012).},
as well as according to their photometric properties. However, a long-term goal of SN classification is to associate observationally-defined 
classes with specific physical explosive phenomena. We show here that this aspiration is now finally coming to fruition, and we establish
the SN classification scheme upon direct observational evidence connecting SN groups with specific progenitor stars. Observationally,   
the broad class of Type II SNe ($\S~\ref{sec:TypeII}$) contains objects showing strong spectroscopic 
signatures of hydrogen, while objects lacking such signatures are of Type I, which is further divided to numerous subclasses ($\S~\ref{sec:TypeIa}, \S~\ref{sec:TypeIbc}$). Recently a class of super-luminous SNe (SLSNe, typically 10 times more luminous than standard events) has been identified, and it is dicussed in $\S~\ref{sec:SLSN}$. We end this chapter by
briefly describing in $\S~\ref{sec:new}$ a proposed alternative classification scheme that is inspired by the stellar classification system. This system presents our emerging physical understanding of SN explosions, while clearly separating robust observational properties from physical inferences that can
be debated. This new system is quantitative, and naturally deals with events distributed along a continuum, rather than being strictly 
divided into discrete classes.Thus, it  
may be more suitable to the coming era where SN numbers will quickly expand from a few thousands to millions of events. }

\section{Introduction: the physical basis of supernova classification}
\label{sec:intro}

The first classification of SN explosions was introduced by Minkowski (1941) based on spectroscopic observations of 14 events. Minkowski designated
the larger subgroup of 9 nearly homogeneous events as Type I, while the other 5 events were designated as Type II. The spectra of Type II SNe were
quite diverse but all showed signatures of hydrogen, while Type I SNe did not. Wheeler \& Levreault (1985) identified a sub-class of SNe I which differs
spectroscopically from the dominant population, and Elias et al. (1985) coined the terms SN Ia for the dominant group and SN Ib for this new subclass.
Harkness et al. (1987) identified signatures of He I in peak spectra of SNe Ib and Wheeler \& Harkness (1990) introduced the term SN Ic for a subclass
of SNe Ib that did not show strong helium but were otherwise similar to SNe Ib (and different than SNe Ia). We will provide positive definitions for these
classes below. Among Type II SNe, Barbon et al. (1979) divided these into two photometric subclasses characterised by light curves showing prominent
plateaus (II-P) or declining in a linear fashion (II-L). Two additional spectroscopic subclasses of SNe II were introduced later: SNe IIb that transition
from having hydrogen-rich early spectra to He-dominated SN Ib-like events near peak (Filippenko 1988) and SNe IIn that show strong and relatively 
narrow emission lines of hydrogen (Schlegel 1990). Filippenko (1997) reviewed SN classification in detail. Since 1997, numerous additional SN groups have been identified and new sub-classes proposed, most notably the class of broad-line SNe Ic (SN Ic-BL) that are associated with high-energy
Gamma-Ray Bursts (GRBs) and X-Ray Flashes (XRFs; Woosley \& Bloom 2006) and the newly-defined class of superluminous supernovae (SLSN; Gal-Yam 2012); all of these will be reviewed in detail below. However, it is now feasible to reformulate the classification of SNe based on their physical origin, rather than view it as a histroical sequence of large classes being split into smaller groups based on certain 
observed features. We will attempt to do this now.

Table~\ref{tab:phys-classes} summarizes the information we have regarding the physical origin of the main classes of SNe. In general, SN explosions
are theorised to result from one of two broad variants of models: the thermonuclear explosion of white dwarf stars or explosive phenomena that follow the terminal collapse of the core of a massive star, see section 6 of this handbook, ``Explosion Mechanisms of Supernovae", for details.
We now have direct and strong evidence associating normal SNe Ia with white dwarf explosions, and many of the other classes with massive star
explosions. Some ambiguity remains for some sub-classes of SNe Ia, Ib, and Ic for which only circumstantial evidence, such as host galaxy population
and SN location studies exist, which provide similar results for these SN I sub-classes (e.g., Perets et al. 2010). A similar situation exists for a sub-class of interacting SNe that are technically members of 
the SN IIn class, but are often assumed to result from white dwarf explosions that interact with circumstantial material (SNe Ia-CSM, e.g., Hamuy et al. 2003, Silverman et al. 2013a; see $\S$~\ref{subsec:TypeIa:pec}). As the most massive white dwarf stars arise from progenitor stars with initial masses just below that of stars the end their life in core-collapse explosions, this is not unexpected.

\begin{table}
\caption{Major supernova classes and their physical origin}
\label{tab:phys-classes}       
\begin{tabular}{p{1.3cm}p{2.4cm}p{4.0cm}p{4.0cm}}
\hline\noalign{\smallskip}
Class & Physical origin & Evidence & Notes \\
\noalign{\smallskip}\svhline\noalign{\smallskip}
SN Ia & White Dwarf  & Size of SN 2011fe progenitor & Nugent et al. 2011; Bloom et al. 2012\\
          &                       & Host galaxy population & For the regular and SN1991bg subclasses; Perets et al. (2010)\\
\noalign{\smallskip}
SN Ib & Massive star  & Shock breakout from SN 2008D & Soderberg et al. (2008)\\
          &                       & Apparent massive progenitor for iPTF13bvn & Cao et al. (2013)\\
\noalign{\smallskip}
SN Ic  & Massive star  & Host galaxy population and locations & Direct evidence still missing\\
\noalign{\smallskip}
SN Ic-BL & Massive star & Association with GRBs/XRFs & Woosley \& Bloom (2006)\\
          &                       & Shock breakout from SN 2006aj & Waxman et al. (2007), Nakar (2015)\\
\noalign{\smallskip}
SN II  & Massive star & Direct progenitor identifications  & Conclusive evidence for II-P and SN 1987A-like subclasses; likely also for II-L (Smartt 2009)\\
\noalign{\smallskip}
SN IIb  & Massive star & Direct progenitor identifications & Maund et al. (2004), Smartt (2009)\\
\noalign{\smallskip}
SN IIn  & Massive star & Direct progenitor identifications & Gal-Yam \& Leonard (2009), Smartt (2009)\\
\noalign{\smallskip}
SLSN  & Massive star & Light curve and host galaxy properties & Direct evidence still missing\\
\noalign{\smallskip}\hline\noalign{\smallskip}
\end{tabular}
\end{table}

\begin{figure}[ht]
\sidecaption[t]
\includegraphics[width=120mm,trim=50 190 50 190 mm, clip=true]{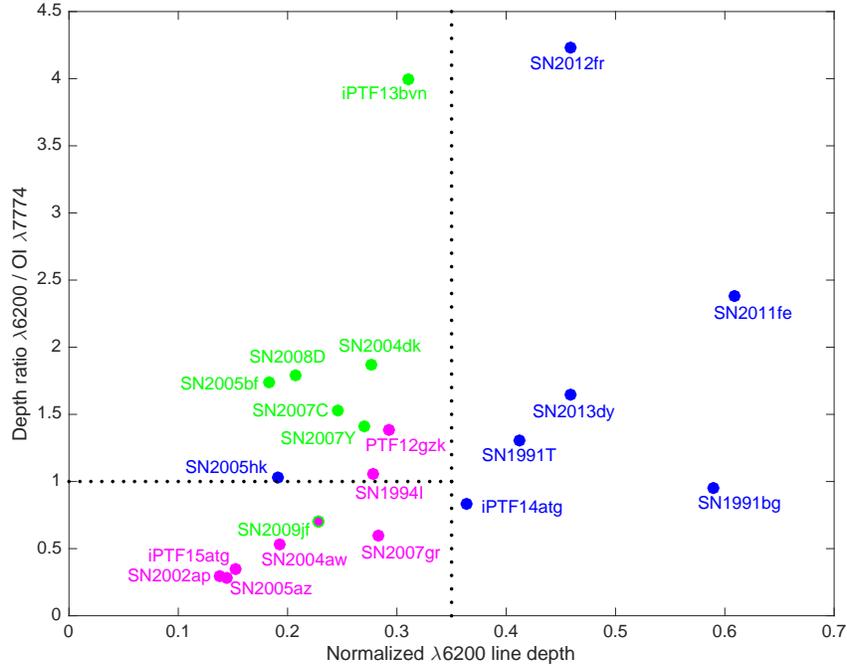}
%
\caption{A classification diagram for Type I SNe based on relative absorption line depth values in peak spectra. Line depths are defined as the relative absorption with respect to the pseudo-continuum, following, e.g., Silverman et al. (2012). Solid circles are measured values from events shown in Fig.~\ref{Fig_Ia_peak}, Fig.~\ref{Fig_Ib_peak}, Fig.~\ref{Fig_Ic_peak} and Fig.~\ref{Fig_Ic_BL_peak}, as well as several other well-measured events from Valenti et al. (2011; SN 2011jf), Stritzinger et al. (2009; SN 2007Y), Gal-Yam et al. (2016, in preparation; SN 2004dk, SN 2005az), Liu et al. (2016; SN 2007C), Benetti et al. (2011; SN 1999dn), Ben-Ami et al. (2012; PTF12gzk), Taddia et al. (2016b, iPTF15dtg), Cao et al. (2015; iPTF14atg), Pan et al. (2015; SN 2013dy), Childress et al. (2013; SN 2012fr), and Silverman et al. (2012; SN 2005hk). Normal and SN 1991bg-like events show deep Si II absorption (depth $>0.35$) around $\lambda6150$\AA. SNe Ib (green) generally show much shallower OI $\lambda 7774$\AA\ absorption relative to the feature around $\lambda6150$\AA\ (depth ratio $>1$) when compared with SNe Ic (megenta). The peculiar SN 2002cx-like subclass of SNe Ia (SN 2005hk shown) have similar locations to SNe Ib in this diagram, but are not similar to most SNe Ic. Some ambiguity between SNe Ib and SNe Ic may exist around depth ratio of $\sim1$; the unusual Type Ib SN 2009jf and Type Ic PTF 12gzk stand out, see discussion in text.}
\label{pEWfig}       
\end{figure}

All major SN classes can be defined based on the peak spectral properties. Normal Type Ia SNe show
strong signatures of Si II near peak (Fig.~\ref{Fig_Ia_peak}) with relative line depth values of a(Si II $\lambda6150$\AA\,)$>0.35$, (following the notation of Silverman et al. 2012; Fig.~\ref{pEWfig}). They also show prominent S II absorption lines that are not seen in other SN classes (Fig.~\ref{Fig_Ia_peak}). The common Type Ia SN 1991bg-like subclass shares these spectral charateristics. There is strong evidence (Table~\ref{tab:phys-classes}) that both normal and SN 1991bg-like SNe Ia result from white dwarf explosions.

SNe Ib are defined by displaying prominent He I $\lambda\lambda\lambda5876$\AA\,6678\AA\,7065\AA\, absorption lines in near-peak spectra (Fig.~\ref{Fig_Ib_peak}), as well as relatively shallow OI $\lambda7774$\AA\, lines (Fig.~\ref{pEWfig}; see e.g., Matheson et al. 2001; Liu et al. 2016). These events also show an absorption line near $\lambda6150$\AA\, whose nature is debated. Following Parrent et al. (2015) and Liu et al. (2016) it seems like the evidence prefers associating this line with hydrogen H$\alpha$ over Si II (contrary to, e.g. Filippenko 1997). There is strong evidence (Table~\ref{tab:phys-classes}) connecting this class with massive star progenitors, which seem to have all retained only a very small (but non-zero)  fraction of their original hydrogen envelope. 

Normal SNe Ic are overall similar to SNe Ib (Fig.~\ref{Fig_Ib_peak}) but (by definition) do not show the three characteristic He I features near peak. However, they often do show a prominent $\lambda6150$\AA\, feature, that we associate with Si II (rather than with hydrogen, but see below), as well as a strong OI $\lambda7774$\AA\,. As shown in Fig.~\ref{pEWfig}, the location of these events in the line depth ratio diagram allows one to differentiate SNe Ic from both SNe Ia (a($\lambda6150$)$<35$) as well as from SNe Ib (a($\lambda6150$\AA\,)/a(OI $\lambda7774$\AA\,)$<1$), regardless of the detection or lack of He I lines. While their similarity to SNe Ib and environmental studies suggest these explosions result from massive stars, conclusive evidence at the level existing for most other SN classes is still lacking.

The subclass of broad-line SNe Ic (SN Ic-BL, Fig.~\ref{Fig_Ic_BL_peak}) show extreme line blending at peak, indicating emission from material moving at a wide range of velocities extending to 0.1c and beyond. A feature that can be identified with some certainty is the $\lambda6150$\AA\,feature (Fig.~\ref{Fig_Ic_BL_peak}) while the IR Ca II and OI $\lambda7774$\AA\, blend into a single extremely broad feature for the most extreme events. This group of events
attracted a lot of attention due to the association of some of these events with high-energy GRBs and XRFs, (e.g., Woosley \& Bloom 2006), and there is strong evidence that SNe Ic-BL arise from massive star progenitors. 

We will refer to SNe whose photospheric spectra are dominated by strong, broad hydrogen lines as spectroscopically regular SNe II (Fig.~\ref{Fig_II_regular_peak}). This group includes the II-P and II-L groups mentioned above, as well as slowly-rising events similar to SN 1987A (e.g., Taddia et al. 2016a and references therein); the separation of this group into additional subclasses is further discussed below and in Chapter 3.2. There is strong evidence that regular SNe II result from the explosion of massive, supergiant stars (Smartt 2009 and references therein).  

SNe whose spectra are initially dominated by hydrogen, but later develop strong He I lines (resembling SNe Ib) belong to the class of Type IIb SNe (Fig.~\ref{Fig_IIb_peak}). The strength and persistence of hydrogen lines in these events varies. Recent works indicating that most and perhaps all SNe Ib also show traces of hydrogen in their spectra (e.g., Liu et al. 2016; Parrent et al. 2016; Fig.~\ref{Fig_IIb_peak}) may actually suggest that the Ib and IIb classes should be unified into a single class (IIb), with the designation ``SN Ib'' saved only to pure, hydrogen-free events; it is not clear whether such events have yet been observed. In any case, there is strong evidence that SNe IIb arise from the explosion of massive stars; in at least some cases these progenitors are yellow or orange supergiants (i.e., hotter than typical red supergiants; Smartt 2009).    

Another spectroscopically-defined subclass of hydrogen-rich events are Type IIn SNe, that show Balmer emission lines with often complex profiles, including narrow ($<$ few hundreds km\,s$^{-1}$), intermediate-width ($\sim2000$\,km\,s$^{-1}$) and sometimes broader components (e.g., Schlegel 1990; Filippenko 1997; Kiewe et al. 2012; Fig.~\ref{Fig_IIn_peak}). The narrow and intermediate-width emission lines are often assumed to result from the SN ejecta interacting with slowly-moving massive CSM (e.g., Chugai \& Danziger 1994; Smith 2014). This class is diverse, but at least some of these explosions arise
from massive progenitors (e.g., Gal-Yam et al. 2007; Gal-Yam \& Leonard 2009; Smith et al. 2011).
The SN IIn class is likely contaminated by other types of explosions, for example, Ia-CSM events resulting from white dwarf explosions within a hydrogen-rich CSM (e.g., Hamuy et al. 2003; Silverman et al. 2013a; $\S$~\ref{subsubsec:Ia:pec:CSM}). 

The last major class of SNe to be added to the growing list is superluminous SNe (SLSNe; reviewed by Gal-Yam 2012). These events are fiducially defined as being brighter than $-21$\,mag at peak, but more physical definitions are being sought. While no strong direct evidence are yet in hand, it is most likely that SLSNe result from massive star explosions; see below and in Chapter 3.8. 
 
\section{Type Ia Supernovae}
\label{sec:TypeIa}

Type Ia SNe are the most commonly observed class of objects. The most luminous of all common
SN types, they are strongly over-represented in flux-limited surveys, accounting, for example, for
$67\%$ of the spectroscopically-confirmed SNe discovered by the PTF and iPTF surveys, and
$60\%$ of the events classified by the large public ESO survey PESSTO (Smartt et al. 2015). Using
the newly-launched automated Transient Name Server (TNS\footnote{http://wis-tns.weizmann.ac.il/})
we find a similar fraction of the SNe designated during 2016 so far ($60\%$ of the total number and $65\%$ of the SNe brighter than 19\,mag) are spectroscopically SNe Ia. 

\subsection{Regular Type Ia Supernovae}
\label{subsec:TypeIa:reg}

\begin{figure}[ht]
\sidecaption[t]
\includegraphics[width=120mm,trim=0 180 0 170 mm, clip=true]{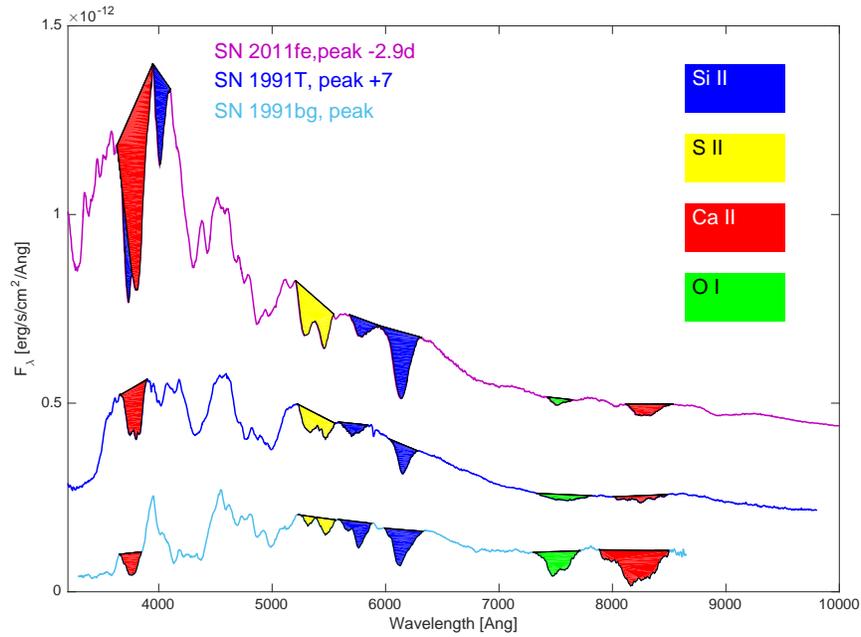}
%
\caption{A spectrum of SN 2011fe (top), a normal Type Ia supernova, obtained 2.9 days prior to peak B-band magnitude. This Hubble Space Telescope 
spectrum from Mazzali et al. (2014) extends further into both the UV and the IR than shown, and is not compromised by Telluric absorption correction
residuals. The spectrum is dominated by absorption lines from
intermediate-mass elements, major features are marked. Additional absorption features are due to iron-group elements, for additional details see
Mazzali et al. (2014). Events belonging to the SN 1991T-like subclass have similar spectra to normal
events after peak (middle; spectrum from Filippenko et al. 1992a), but show very distinctive features earlier (Fig.~\ref{Fig_Ia_early}). Members of the SN 1991bg-like class (bottom spectrum from Filippenko et al. 1992b) are also quite similar, but show a distinctive broad absorption trough
(due to blends of Fe-group elements, e.g., Mazzali et al. 1997) spanning $4000-4500$\,\AA\ at peak.}
\label{Fig_Ia_peak}       
\end{figure}

Fig.~\ref{Fig_Ia_peak} shows the peak optical spectrum of the nearby, spectroscopically normal SN 2011fe. 
The spectrum does not show any lines of hydrogen or helium. It is dominated by absorption lines from intermediate-mass elements, and in particular 
it shows a very prominent Si II absorption around $6100$\AA, as well as lines from O, S, Ca and blends of Fe-group elements. Looking at Fig.~\ref{pEWfig}, we can see that normal SNe Ia like SN 2011fe are distinguished from all other types of H-poor events by having Si II lines that are both deep and dominant over O aborption lines (large Si/O depth ratio), uniquely placing these objects in the upper right corner of Fig.~\ref{pEWfig}. The typical expansion 
velocities of normal SNe Ia as deduced from the minimum of the Si II lines at peak are around 
$11000$\,km\,s$^{-1}$ (e.g., Silverman et al. 2012; Maguire et al. 2014), and the typical peak magnitudes 
are M$_B=-19.1$\,mag (Ashall et al. 2016) and M$_R=-18.67$\,mag (Li et al. 2011) . Early spectra of 
SNe Ia are shown in Fig.~\ref{Fig_Ia_early} while late nebular spectra are shown in Fig.~\ref{Fig_Ia_late}.
The properties of normal SNe Ia are reviewed in Chapter 3.4. 

\subsection{Peculiar Type Ia Supernovae}
\label{subsec:TypeIa:pec}

The spectroscopic homogeneity of normal SNe Ia serves to highlight unusual or peculiar events,
that are more notable in this context. Here we review the more commonly observed sub-classes of
peculiar SNe Ia. A more general description of this population is provided in Chapters 3.5 and 3.6. \\
  
\subsubsection{SN1991T-like events}

The prototype of this class of unusual events was pointed out by Filippenko et al. (1992a) as an unusual SN with a distinctive pre-peak spectrum dominated by Fe III lines (Ruiz-Lapuente et al. 1992; Mazzali et al. 1995) that later evolves to resemble normal SNe Ia, as well as being more luminous (by $\sim0.5$\,mag) than regular SNe Ia. Analysis by Mazzali et al. (1995) showed this event was hotter and more luminous than regular events due to the synthesis of a larger amount of radioactive $^{56}$Ni in the explosion, but is
otherwise similar to normal SNe Ia. Silverman et al. (2012) and Blondin et al. (2012), show that the expansion velocities of SN 1991T-like events, as measured from the Si II lines at peak, are consistent, within the errors, with those of regular SNe Ia. 
Li et al. (2001) find that up to $20\%$ of SNe Ia belong to this
sub-class, so these are not rare events.

\begin{figure}[ht]
\sidecaption[t]
\includegraphics[width=120mm,trim=50 200 50 200 mm, clip=true]{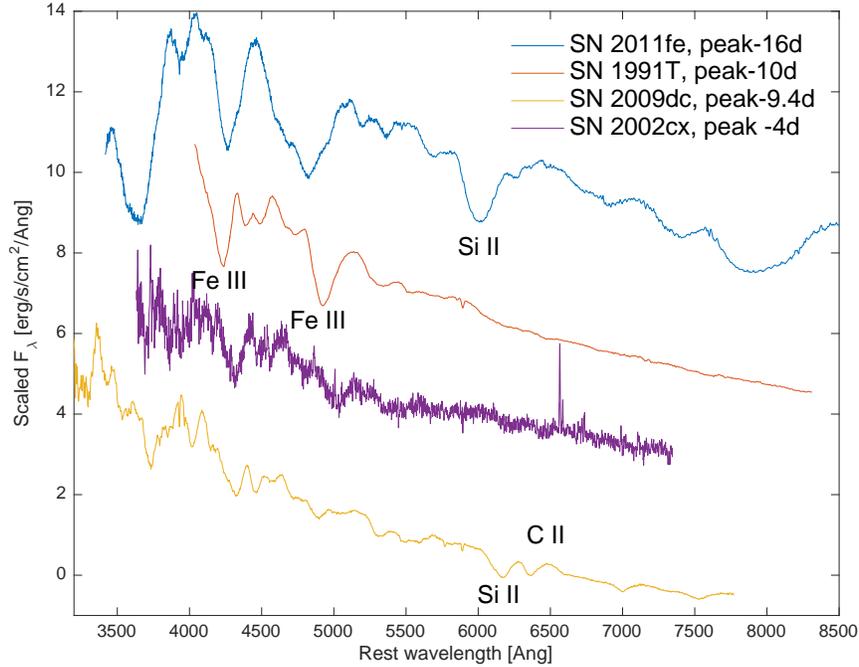}
\caption{Early spectra of SNe Ia. The normal SN 2011fe was discovered very shortly after explosion;
a spectrum at -16d prior to peak from Nugent et al. (2011; top) is shown. Spectra of SN 1991T prior to peak (a spectrum at -10d from Filippenko et al. 1992 is shown) lack strong Si II features due
to its high photospheric temperature; instead, prominent Fe III lines are seen. The earliest spectrum of 
SN 2002cx (at -4d, from Li et al. 2003) has similar Fe III features but at much lower velocities. An early spectrum of the ``Super-Chandra'' event SN 2009dc (-9.4d, from Taubenberger et al. 2011) shows weak and low-velocity Si II and prominent CII.}
\label{Fig_Ia_early}       
\end{figure}
 
\begin{figure}[ht]
\sidecaption[t]
\includegraphics[width=120mm,trim=50 200 50 200 mm, clip=true]{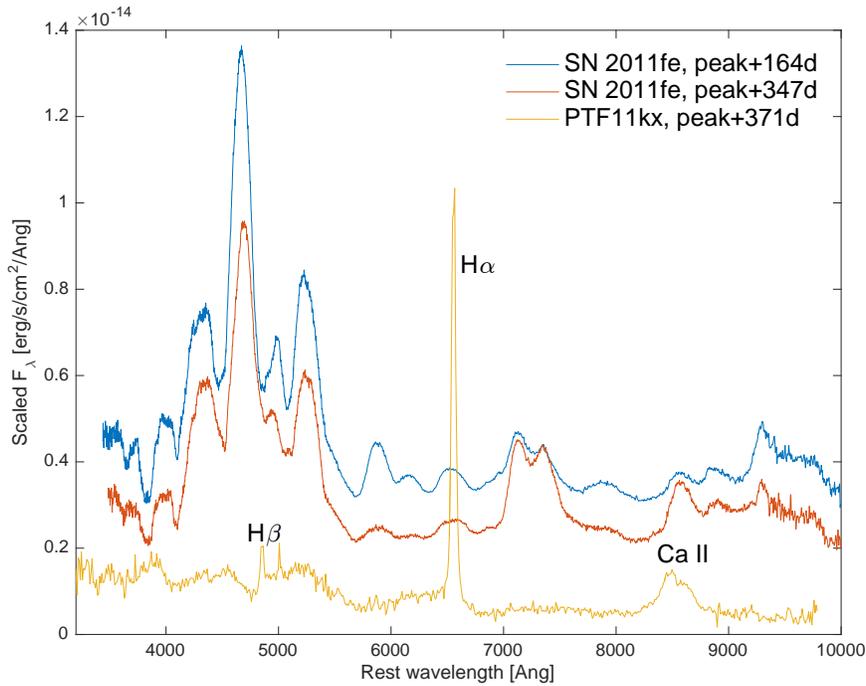}
\caption{Late-time (nebular) spectra of the normal Type Ia SN 2011fe (from Mazzali et al. 2015) 
are dominated by strong lines of Fe-group elements (see, e.g., Mazzali et al. 2015 for line identifictions).
In contrast, the late-time spectra of the SN Ia-CSM PTF11kx (Silverman et al. 2013b) are 
dominated by strong hydrogen Balmer lines, likely from the shocked CSM.
}
\label{Fig_Ia_late}       
\end{figure}

\subsubsection{SN1991bg-like events}

The prototype of this class of peculiar SNe Ia was described by Filippenko et al. (1992b; see also
Ruiz-Lapuente et al. 1993). This
class of events are substantially less luminous than normal SNe Ia (typically by $R\sim1.1$\,mag at peak; Li et al. 2011).
Their spectra resemble those of normal SNe Ia (Fig.~\ref{Fig_Ia_peak}) except for a prominent
broad absorption trough at $4000-4500$\,\AA\, which results from a blend of Fe-group elements
dominated by Ti II (Filippenko et al. 1992b, Mazzali et al. 1997), and reflects a low photospheric
temperature. The light curves of these events are also quite distinctive as they lack the secondary 
IR maximum seen in normal SNe Ia. Kasliwal et al. (2008) present an extremely faint member of this group, and discuss the light curve properties of this faint class. Howell (2001) inspected the host
galaxies of such objects and showed these indicate old progenitor systems for members of this class.
Silverman et al. (2012) and Blondin et al. (2012), show that the expansion velocities of SN 1991bg-like events, as measured from the Si II lines at peak, are consistent, within the errors, with those of 
regular SNe Ia. 
Li et al. (2001) finds that $16\%$ of SNe Ia belong to this
class, but they tend to be under-represented in large flux-limited surveys due to their intrinsic 
faintness.

\subsubsection{``Super-Chandra'' events}

\begin{figure}[ht]
\sidecaption[t]
\includegraphics[width=120mm,trim=50 200 50 200 mm, clip=true]{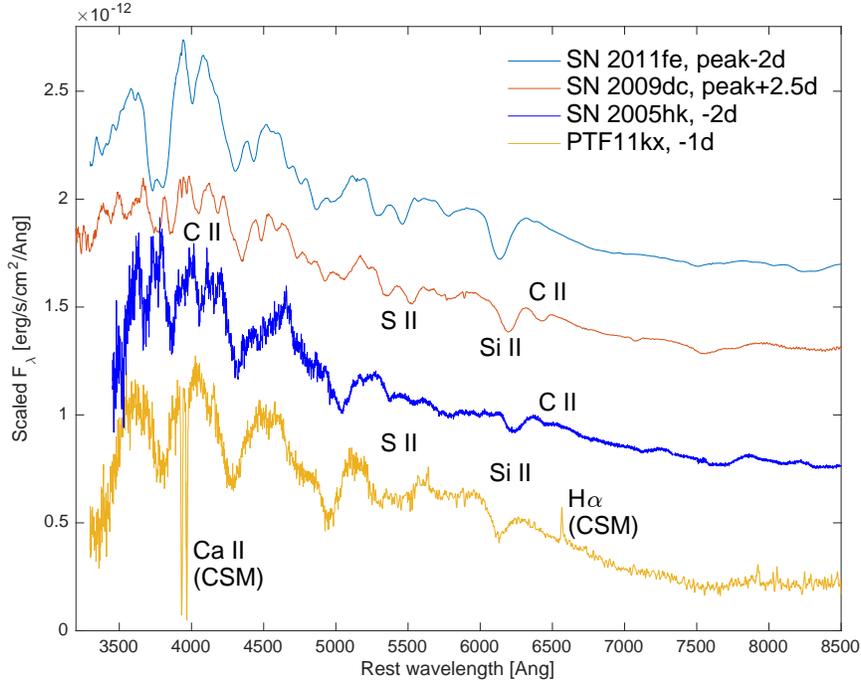}
\caption{Peak spectra of peculiar SNe Ia compared to the normal SN 2011fe (top). The ``Super-Chandra'' event SN 2009dc (from Taubenberger et al. 2011) shows the hallmark Si II and SII lines, though at notably lower velocities, as well as prominent CII lines. The SN 2002cx-like (``Iax'') event SN 2005hk also shows the Si II and SII lines, as well as possibly CII, though at lower contrast and lower
expansion velocities. The spectrum shown is a combination of the high quality Keck spectrum from 
Silverman et al. (2011) with a spectrum extending further to the blue from Blondin et al. (2012), 
obtained on the same night. The Ia-CSM event PTF11kx (spectrum from Dilday et al. 2011) 
shows standard SN Ia features, including Si II and SII, as well as extrenely strong Ca II H+K absorption and H$\alpha$ emission from the surrounding CSM. These CSM features strongly evolve with time.  
}
\label{Fig_Ia_peculiar_peak}       
\end{figure}

Howell et al. (2006) presented the discovery of a very luminous (by about $0.87$\,mag in absolute
V magnitude) SN Ia with low expansion velocities ($8000$\,km\,s$^{-1}$ at peak), which they
interpret as indicating a large total ejecta mass, surpassing the Chandrasekhar mass, leading
to the moniker "Super-Chandra" for these rare objects. A few additional examples of such objects
have been published since, including SN 2006gz (Hicken et al. 2007), SN 2009dc (Tanaka et al. 2010), and SN 2007if (Scalzo et al. 2010). The distinctive spectroscopic features of this class are strong CII lines around peak (Fig.~\ref{Fig_Ia_peculiar_peak}). While very luminous, these events do not show
strong Fe III at early times (Fig.~\ref{Fig_Ia_early}). The small detected number of these events in spite of their luminosity suggest their volumetric rate is low, probably below $1\%$ of the SN Ia rate.

\subsubsection{SN 2002cx-like events}

Li et al. (2003) presented a very peculiar object, SN 2002cx. Early spectra of this object (Fig.~\ref{Fig_Ia_early}) resemble those of the SN 1991T-like variant of SNe Ia, while peak spectra (Fig.~\ref{Fig_Ia_peculiar_peak}) show some similarity to normal SNe Ia, but with notably lower expansion velocities ($<8000$\,km\,s$^{-1}$; Foley et al. 2013, White et al. 2015). Additional members of this group were discovered in later years and sample papers have been presented by Foley et al. (2013; introducing the name ``Iax'' for this class), White et al. (2015) and Foley et al. (2016). The inclusion of events with prominent He I lines in this class of SNe Ia is debated
(Foley et al. 2013; 2016 vs. White et al. 2015) as is the possible association of this class with similar
low-velocity events with cooler spectra (SN 2002es-like events, e.g., Ganeshalingam et al. 2012; White et al. 2015; Cao et al. 2015; 2016). The mean peak $R$-band magnitude of the objects listed by White
et al. (2015) is M$_{R}=-17$\,mag. The rate of these events has been initially estimated to be low (a few \%, Li et al. 2011), but Foley et al. (2013) argue it is much higher, and can be as much as $30\%$ of the 
SN Ia rate. This class is further discussed in Chapter 3.6.

\subsubsection{SNe Ia-CSM}
\label{subsubsec:Ia:pec:CSM}

Hamuy et al. (2003) presented the discovery of SN 2002ic and showed that the spectral data of this supernova could be decomposed into spectra of a peculiar SN Ia of the SN 1991T-like variety, combined with a blue continuum source and strong hydrogen Balmer lines. They thus suggested that this event resulted from the explosion of a SN Ia within an envelope
of circumstellar material (CSM). Following this work additional similar events were discovered, while Benetti et al. (2006) challenged the SN Ia nature of the explosion and suggested instead that these events result from stripped-envelope core-collapse SNe (e.g., SNe Ic) that interact with hydrogen-rich CSM. The discovery of PTF11kx, a nearby event whose early spectra were compellingly similar to those of SNe Ia (Dilday et al. 2012; Fig.~\ref{Fig_Ia_peculiar_peak}), and became dominated by CSM interaction later on (Silverman et al. 2013b; Fig.~\ref{Fig_Ia_late}) established that SNe Ia that interact with
CSM exist. Silverman et al. (2013a) assembled a large sample of candidate SN Ia-CSM events; the
nature of these objects and the level of contamination by SNe IIn remains an open issue (Leloudas et al. 2015; Inserra et al. 2016a). The mean $r$-band peak magnitude of the values listed in Silverman
et al. (2013) is M$_{r}=-20.2$\,mag. Due to the often strong interaction signatures, it is difficult to 
determine the peak expansion velocity of objects of this class.
From the frequency of SN Ia-CSM candidates in the PTF survey (Dilday et al. 2012; Silverman et al. 2013a) we can deduce these are intrinsically rare events, accounting for less than $1\%$ of the SN Ia population.

\subsubsection{Other peculiar SNe Ia} 

Among the large numbers of SNe Ia discovered and studied over the years, several other events have
been pointed out to be peculiar in their spectroscopic and/or photometric properties. These are unique events that may represented even rarer SN Ia subclasses, including SN 2000cx and its twin SN 2013bh (Li et al. 2001; Silverman et al. 2013c), PTF10ops (Maguire et al. 2011), SN 2010lp (Taubenberger et al. 2013) and PTF09dav (Sullivan et al. 2011). Chapter 3.5 further discusses peculiar SNe Ia.

\section{Type Ib and Type Ic  Supernovae}
\label{sec:TypeIbc}

Type Ib and Type Ic SNe do not show strong hydrogen features, setting them apart from hydrogen-rich
SNe II, and differ from regular SNe Ia in having a shallower depth of the $\lambda6150$\AA\ line (Fig.~\ref{pEWfig}). Their classification within the hydrogen-poor Type I class is more subtle, and will be discussed below. 
 
\subsection{Regular Type Ib Supernovae}
\label{subsec:TypeIb:reg}

\begin{figure}[ht]
\sidecaption[t]
\includegraphics[width=120mm,trim=90 225 115 225 mm, clip=true]{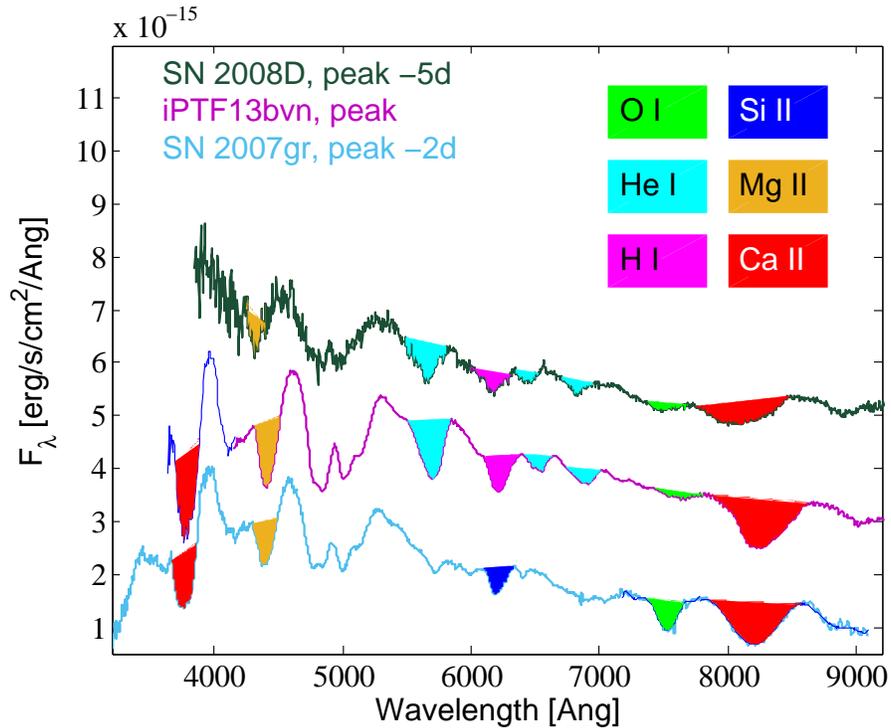}
%
\caption{Spectra of events considered to be regular Type Ib SNe (top; SN 2008D from Modjaz et al. 2009 and iPTF13bvn from Cao et al. 2013) compared with a spectrum of a regular Type Ic SN (bottom; see Fig.~\ref{Fig_Ic_peak}) . Major absorption features are marked, while the spectral shape in the blue part is dominated by multiple absorption features from iron-group elements.}
\label{Fig_Ib_peak}       
\end{figure}

SNe Ib show, by definition, strong He I absorption features around peak (Fig.~\ref{Fig_Ib_peak}). 
For the large majority of observed Type I events the clear detection of He I lines is sufficient to classify these as regular SNe Ib; Fig.~\ref{Fig_Ib_peak} shows the spectral appearance of such events. 

There are two rare classes that complicate this statement though. There are a few SNe with He I in their peak spectra and low expansion velocities. Foley et al. (2013) include these in the class of SN 2002cx-like SNe Ia (``Iax''), and suggest a physical origin in WD progenitor systems. White et al. (2015)  
challenge this association. At this time it is not clear whether the ``Iax'' group is indeed the only subclass of SNe Ia that shows He I in its peak spectra, or whether the He-rich events discussed
by Foley et al. (2013), SN 2007J and SN 2004cs (that may also show hydrogen features, Rajala et al. 2005; White et al. 2015), are unrelated low-velocity SNe Ib/IIb; see also Chapter 3.6.  
 
Another rare class is composed of events that show He I features around peak, but later develop
nebular spectra dominated by Ca II lines. This class was first discussed by Perets et al. (2010) who showed strong evidence that these events arise from an old stellar population and proposed a WD origin for this class. It is clear that regardless of its exact origin, this subclass of events is unlikely to 
result from the same massive-star progenitors as regular SNe Ib; we discuss this group below in $\S~\ref{subsubsec:TypeIb:pec:ca}$.

An important issue related to the classification of SNe Ib and their distinction from SNe Ic and SNe IIb (see below) is the nature of the strong absorption seen around $\lambda6150$\AA\ at peak. Following previous works (e.g., Parrent et al. 2016; Liu et al. 2016) we consider it most likely that this feature is dominated by hydrogen Balmer H$\alpha$ absorption, based on the striking spectral similarity with SNe IIb (Fig.~\ref{Fig_IIb_peak}) where the identification of this feature
is clear. Assuming this feature is H$\alpha$, the hydrogen expansion velocities are higher than those 
measured from He or Fe lines (Liu et al. 2016), as expected for a homologous explosion of a stratified progenitor star; an alternative identification as Si II (with a lower rest-frame wavelength) leads to Si II velocities which sometimes fall below the Fe velocities. Some contribution to this line from Si II is possible, and may be the reason for SNe Ib showing bluer $\lambda6150$\AA\ features than SNe IIb (Fig.~\ref{Fig_IIb_peak}; Liu et al. 2016), reflecting an increased contribution from the bluer Si II line. An additional piece of supporting evidence comes from late-time nebular spectra, where SNe Ib seem to show evidence for small amounts of hydrogen (Fig.~\ref{Fig_I_neb_zoom}). 

The distinction between SNe Ib and SNe IIb has been considered extensively by Liu et al. (2016). They find that while all SNe Ib show evidence for hydrogen, the pseudo-equivalent width of H$\alpha$ at peak is stronger in objects reported in the literature as SNe IIb (pEW=$120\pm 24$\AA)
than those reported as SNe Ib (pEW=$31\pm18$\AA), after
they reclassify some objects, though. The fact that the difference between SNe IIb and SNe Ib may only reflect a continuous distribution of residual hydrogen envelope mass perhaps motivates merging  
these two subclasses into one (which should be called IIb, as these events contain hydrogen and 
are therefore SNe II), and reservation of the class of SN Ib to events that show no hydrogen at all, though whether such events exist remains to be seen. The peak expansion velocities of regular 
SNe Ib (based on He or Fe lines) were found by Liu et al. (2016) to be $9000-10000$\,km\,s$^{-1}$,
while Drout et al. (2011) find a mean $R$-band peak absolute magnitude of M$_{\rm R}=-17.9$\,mag.

\begin{figure}[ht]
\sidecaption[t]
\includegraphics[width=120mm,trim=50 200 50 200 mm, clip=true]{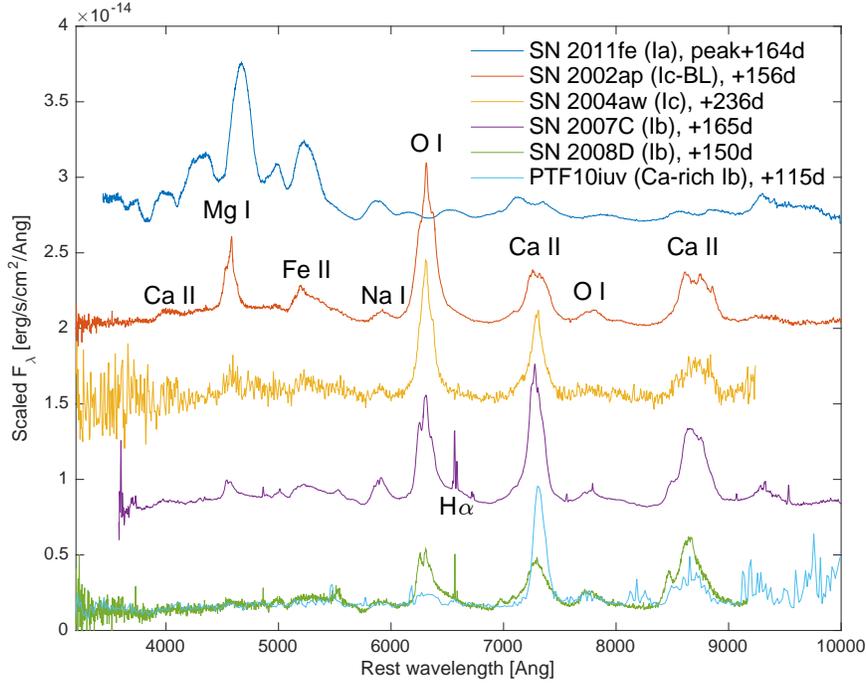}
%
\caption{
Nebular spectra of Type I SNe. The nebular spectra of regular SNe Ia (SN 2011fe from Mazzali et al. 2015 shown) uniquely shows strong blends of Fe-group elements in the blue ($4000-5300$\AA), 
while spectra of SNe Ib (SN 2008D, Modjaz et al. 2009; SN 2007C, Taubenberger et al. 2009) and SNe Ic (SN 2002ap, Foley et al. 2003; SN 2004aw, Taubenberger et al. 2006) are dominated by strong emission of intermediate-mass elements such as O, Mg, and Ca. Ca-rich SNe Ib such as PTF10iuv (Kasliwal et al. 2012) show strong Ca II 
emission and weak or absent OI.
}
\label{Fig_I_neb}       
\end{figure}

\begin{figure}[ht]
\sidecaption[t]
\includegraphics[width=120mm,trim=50 200 50 200 mm, clip=true]{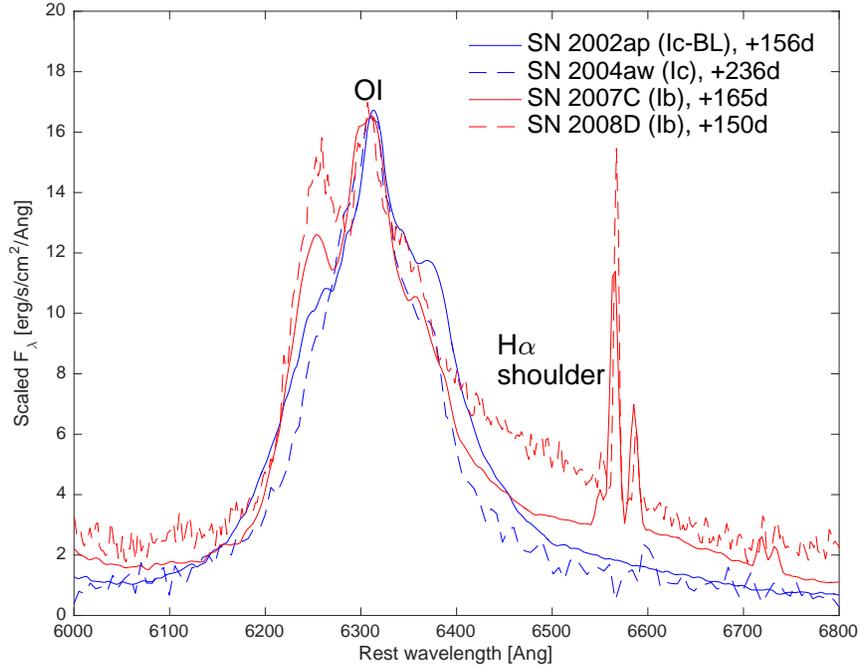}
%
\caption{
Nebular spectra of Type Ib SNe and Type Ic SNe around $6500$\AA\ (see Fig.~\ref{Fig_I_neb} for data sources). Spectra have been normalized at the peak of the strong OI $\lambda6300$\AA\ emission line. Note the pronounced shoulder redwards of this line seen in Type Ib spectra, which is likely a weak contribution from
hydrogen H$\alpha$; this feature is lacking in nebular spectra of SNe Ic. Narrow emission lines (H$\alpha$, NII and SII) are from the host galaxies.}
\label{Fig_I_neb_zoom}       
\end{figure}

\subsection{Unusual Type Ib Supernovae}
\label{subsec:TypeIb:pec}

\subsubsection{Interacting Type Ibn events}
\label{subsubsec:TypeIb:pec:Ibn}

\begin{figure}[ht]
\sidecaption[t]
\includegraphics[width=120mm,trim=50 200 50 200 mm, clip=true]{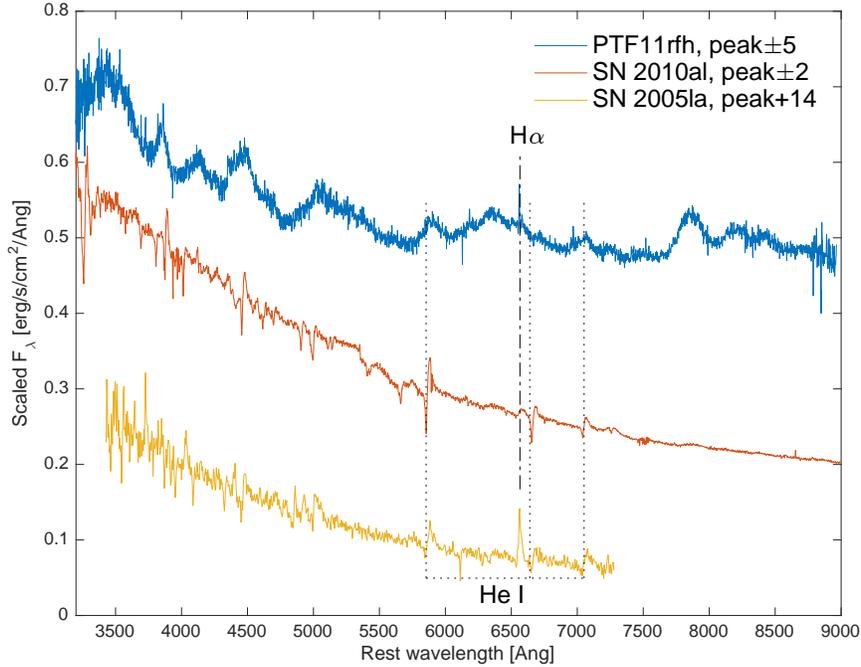}
%
\caption{
Spectra of SNe Ibn obtained shortly after peak. The top two events are representative of the
two classes proposed by Hosseinzadeh et al. (2016) based on width and shape of the He I lines
(dotted). Transitional Ibn/IIn events such as SN 2005la (bottom, Pastorello et al. 2008b) show prominent hydrogen Balmer lines (dash-dot) in addition to He I.}
\label{Fig_Ibn_peak}       
\end{figure}

Type Ibn SNe show strong and relatively narrow  He I emission lines in their early spectra, and were
thus named ``Ibn'', in analogy to SNe IIn, by Pastorello et al. (2007). The class was recognized by Pastorello et al. (2007; 2008a) following the discovery of the nearby event SN 2006jc (Pastorello et al. 2007; Foley et al. 2007). There are $<30$ events of this class known at this time, with recent compliations presented by Pastorello et al. (2016) and Hosseinzadeh et al. (2016). These events can be quite luminous, typically peaking around $-19$\,mag (Hosseinzadeh et al. 2016) and usually
decline quickly after maximum (by $\sim0.1$\,mag\,day$^{-1}$). Some events show double-peaked light
curves (e.g., Gorbikov et al. 2014). Spectroscopically, there is noticeable heterogeneity (Pastorello et al. 2016), including a group of events with prominent hydrogen lines in addition to He I (Pastorello et al.   
2008b). Hosseinzadeh et al. (2016) propose a division into two sub-groups based on early spectra (Fig.~\ref{Fig_Ibn_peak}). The detection of a pre-explosion outburst from the progenitor of SN 2006jc (Pastorello et al. 2007) suggests a massive star origin for these objects.
 
\subsubsection{Double-peaked SNe Ib}
\label{subsubsec:TypeIb:pec:trans}

Folatelli et al. (2006) presented to curious case of SN 2005bf, a very peculiar stripped-envelope SN. This event has a unique light curve with two broad and luminous peaks. Folatelli et al. (2011) claim it was spectroscopically a SN Ic during the first peak and then evolved into a SN Ib during the second 
peak; however, clear detection of He I already in its earlier spectra (Folatelli et al. 2016; Modjaz et al. 2014) indicates it is really a SN Ib (albeit with unusually low He I expansion velocities of $7000$\,km\,s$^{-1}$ or less). Tominaga et al. (2005), Maeda et al. (2007) and Maund et al. (2007) present additional observations and analysis of this unique event. Recently, Taddia et al. (2016, in preparation) identified a possible second member of this rare group.

\begin{figure}[ht]
\sidecaption[t]
\includegraphics[width=120mm,trim=50 145 50 150 mm, clip=true]{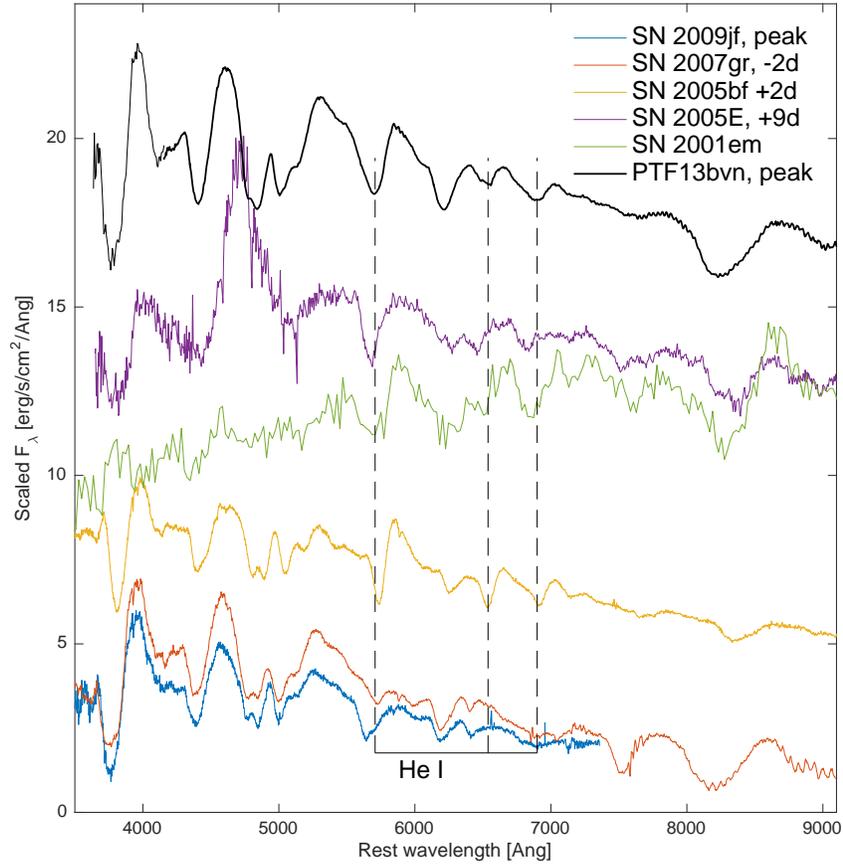}
%
\caption{
Spectra of peculiar SNe Ib around peak, compared with the regular iPTF13bvn (top). SN 2005bf (Folatelli et al. 2006) clearly shows He I lines at somewhat lower expansion velocities, while SN 2005E (Perets et al. 2010) shows the same lines at somewhat higher velocities. A spectrum of SN 2001em (Filippenko \& Chornock 2001; rebinned to increase the S/N) also clearly shows He lines. On the other hand, SN 2009jf (Valenti et al. 2011) does not show clear absorption from both He I $\lambda6678$\AA\ and $\lambda7065$\AA\, and should probably be classified as a SN Ic (compare with Type Ic SN 2007gr; Valenti et al. 2009). }
\label{Fig_Ib_peculiar_peak}       
\end{figure}

\subsubsection{Ca-rich SNe Ib}
\label{subsubsec:TypeIb:pec:ca}

Perets et al. (2010) presented a group of very peculiar events, exemplified by SN 2005E, that was a subclass of SNe Ib based on peak spectra (Fig.~\ref{Fig_Ib_peculiar_peak}), but later developed peculiar nebular spectra dominated by Ca II lines (instead of OI, e.g., Fig.~\ref{Fig_I_neb}) earning these objects the moniker Ca-rich SNe Ib. Perets et al. (2010) presented strong evidence that these 
events did not arise from massive, young-lived progenitors similar to those of regular SNe Ib, and 
suggested instead an origin in binary WD systems. Kasliwal et al. (2012) presented an additional 
compilation of these events, and focussed on the unusual remote galactic locations they seemed to preferntially occur in. Ca-rich events have similar expansion velocities to those of regular SNe Ib, and fainter peaks magnitudes by $\sim2$\,mag, as well as more rapidly declining light curves (Kasliwal et al. 2012). 
While it seems clear that these events are not regular SNe Ib, the exact nature of the progenitor 
systems remains an open question.
  
\subsection{Regular Type Ic Supernovae}
\label{subsec:TypeIc:reg}

\begin{figure}[ht]
\sidecaption[t]
\includegraphics[width=120mm,trim=90 225 115 225 mm, clip=true]{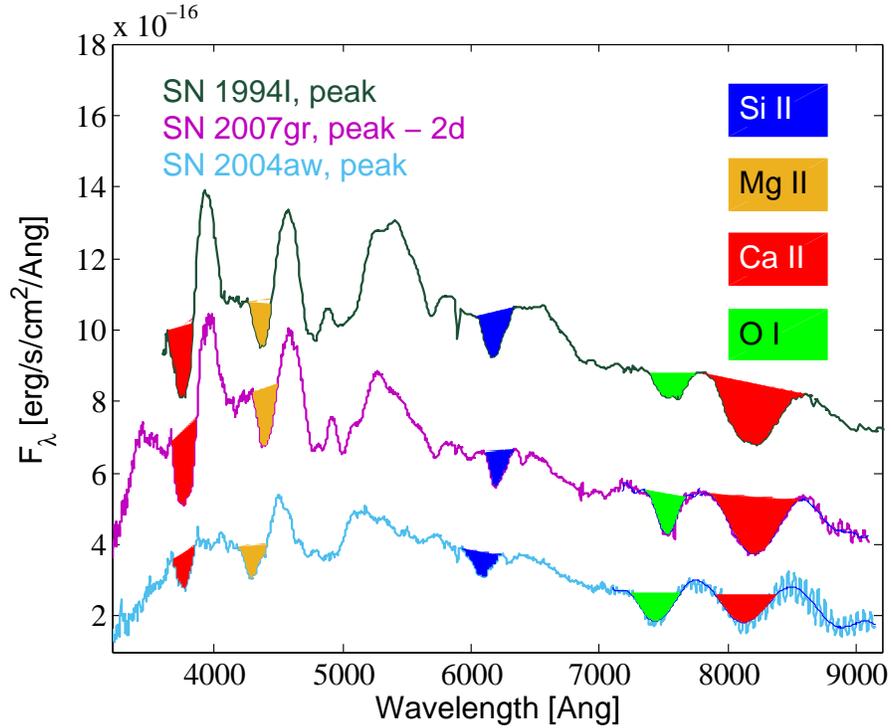}
%
\caption{Spectra of events considered to be regular Type Ic SNe. Major absorption features by intermediate-mass elements are marked, while the spectral shape in the blue part is dominated by multiple absorption features from iron-group elements. Si II absorption is detected, its strength with respect
to other elements (most notably OI) is distinctively weaker than in SNe Ia (Fig.~\ref{pEWfig}).
The well studied SN 1994I in M51 (spectrum from Clocchiatti et al. 1996; dereddened by E$_{\rm B-V}=0.2$\,Mag) is spectroscopically
normal, but photometrically it is an extreme event, e.g., in terms of its rapid decline from peak (Drout et al. 2011; Bianco et al. 2014). SN 2007gr in notable for a conspicous absorption feature redwards of the Si II feature that is likely due to C II $\lambda6580$\AA~ (Valenti et al. 2007). SN 2004aw (Taubenberger et al. 2006) shows higher velocities and more line blending compared to the other events shown, and may be an intermediate event between regular SNe Ic and the broad-line variety (Fig.~\ref{Fig_Ic_BL_peak}).}
\label{Fig_Ic_peak}       
\end{figure}

Unlike SNe Ib and SNe II, Type Ic SNe were not
defined so far by the detection of a certain feature but by the absence of strong features of hydrogen and helium (Filippenko 1997; Fig.~\ref{Fig_Ic_peak}). This negative definition begs the question, whether all objects that are not SNe Ib or SNe II are members of the Type Ic class? In particular, if we claim (Table~\ref{tab:phys-classes}) that SNe Ia and SNe Ic arise from physically different progenitor systems (white dwarfs vs. massive stars), can we point out
positive features that SNe Ic possess that set them apart from SNe Ia? 

The answer is yes. Fig.~\ref{pEWfig} shows that SNe Ic differ from regular SNe Ia at peak, having shallower absorption lines
around $\lambda6150$\AA\ (the nature of which will be discussed below), and in particular, these
lines are sub-dominant to strong $\lambda7774$\AA\ OI lines, placing regular SNe Ic at the lower
left-hand corner of Fig.~\ref{pEWfig}. Another clear observational distinction between Type Ic SNe and SNe Ia is the appearance of the late-time nebular spectra (Fig.~\ref{Fig_I_neb}). This distinction is
clear, but such nebular spectra are observationally difficult to obtain. While regular SNe Ia and regular SNe Ic can be distinguished using peak light spectra, the same is not necessarily true for the many peculiar subclasses of SNe Ia and SNe Ic. We return to this question below. 

The nature of the $\lambda6150$\AA\ absorption in SNe Ic is subject to debate. This feature is clearly 
due to hydrogen in SNe II (and in particular IIb; Fig.~\ref{Fig_IIb_peak}). If we accept that this
feature is dominated by hydrogen also in SNe Ib (Parrent et al. 2016; Liu et al. 2016; $\S$~\ref{subsec:TypeIb:reg}), the similarity between SN Ib and SN Ic peak spectra (Fig.~\ref{Fig_Ib_peak}) may suggest this line is also dominated by hydrogen in spectra of regular SNe Ic (Parrent et al. 2016).
An appealing idea would be that SNe Ib and SNe Ic actually arise from physically similar progenitors, and differ only in the post-supernova photospheric conditions, that are conductive to the formation of He lines in SNe Ib, but not in SNe Ic (e.g., due to weaker radioactive $^{56}$Ni mixing; Dessart et al. 2012). However,
detailed calculations by Hachinger et al. (2012) indicate SNe Ic are genuinely helium and hydrogen 
poor, as do early-time observations by Taddia et al. (2014). We follow therefore Filippenko (1997) and Liu et al. (2016) in assuming that the 
$\lambda6150$\AA\ absorption in SNe Ic is dominated by Si II. The fact that SNe Ic are hydrogen poor
compared to SNe IIb and Ib is further supported by nebular spectra (Fig.~\ref{Fig_I_neb_zoom}), while
the different ratio of $\lambda6150$\AA\ absorption depth to $\lambda7774$\AA\ OI (Fig.~\ref{pEWfig}) further suggests that SNe Ic differ from SNe Ib not only in the lack of He I lines, but also in other
observed properties. 

We thus conclude that we can use Fig.~\ref{pEWfig} to positively define spectroscopically regular SNe Ic
using their peak spectra, and that these events clearly differ from SNe Ia as well as from SNe Ib in properties other than the strength of He I lines. Modjaz et al. (2016) measure the mean expansion velocity at peak of regular SNe Ic to be $9700\pm1600$\,km\,s$^{-1}$, while Drout et al. (2011) find
a mean $R$-band peak magnitude of M$_{\rm R}=-18.3$\,mag. 

\subsection{Broad-line Type Ic Supernovae (Ic-BL)}
\label{subsec:TypeIc:BL}

\begin{figure}[ht]
\sidecaption[t]
\includegraphics[width=120mm,trim=60 200 60 200 mm, clip=true]{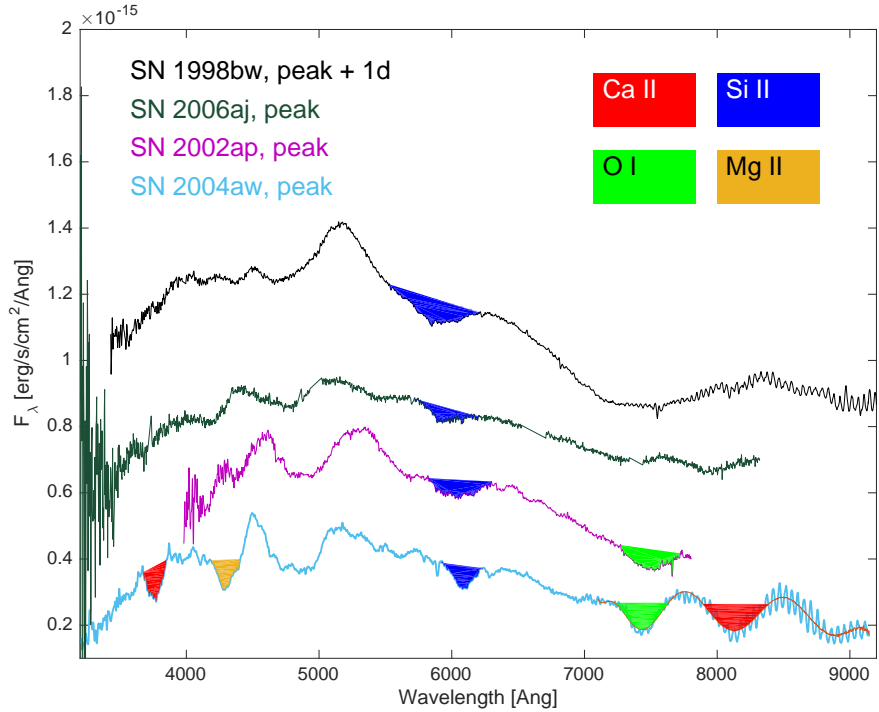}
%
\caption{A series of spectra of Type Ic SNe extending from SN 2004aw, considered a transitional event between normal and broad-line (BL) events (Taubenberger et al. 2006; Fig.~\ref{Fig_Ic_peak}), through the relatively low-energy SN 2002ap (Mazzali et al. 2002; spectrum from Gal-Yam et al. 2002), SN 2006aj, associated with an X-Ray Flash (XRF; Campana et al. 2006, spectrum from Pian et al. 2006) to the energetic SN 1998bw associated with GRB 980425 (Galama et al. 1998, spectrum from Patat et al. 2001). All spectra are around B-band peak. Note the gradual evolution from bottom to top as lines shift to higher velocities and blend together. This sequence establishes a spectral connection between the Ic-BL class and normal SNe Ic. A feature dominated by Si II is marked in all spectra, while other distinct features that are evident in normal SNe Ic (Fig.~\ref{Fig_Ic_peak}) blend together, for example the strong Ca II and OI features seen toward the red.}
\label{Fig_Ic_BL_peak}       
\end{figure}

This subclass of broad-line SNe Ic came into dramatic focus with the discovery of the first well-studied example,
SN 1998bw, in the error region of the Gamma-Ray Burst GRB980425 (Galama et al. 1998). Events
of this subclass remain the only ones to be securely associated with GRBs (e.g., 
Woosley \& Bloom 2006). The
discovery of the nearby SN 2002ap (Gal-Yam et al. 2002; Mazzali et al. 2002; Foley et al. 2003) that
was not associated with a GRB, further boosted the study of this class as a variant of SNe Ic, based on the strong spectroscopic resemblance (Fig.~\ref{Fig_Ic_BL_peak}).   

Broad-line SNe Ic (Ic-BL) are spectroscopically distinct in showing extreme values of velocity 
dispersion at and before peak (Fig.~\ref{Fig_Ic_BL_peak}) that cause individual spectral lines to 
blend together. Modjaz et al. (2016, their Fig. 7) carefully studied the separate effects of large velocity
dispersion that drives line blending (that they model by the velocity width of a convolution kernel required to match the mean regular SN Ic spectra to Ic-BL events) and the effects of  
high expansion velocities (that cause a shift in the minima of spectral lines). Based on this work 
we adopt their definition of SNe Ic-BL as having broad (rather than blue-shifted) absorption lines,
with a useful (if somewhat arbitrary) convolution kernel value of $2000$\,km\,s$^{-1}$. As noted
by Modjaz et al. (2016), velocity dispersions and expansion velocities are strongly correlated, but 
the existence of outliers such as PTF10vgv (Corsi et al. 2011) that has broad lines (high dispersion) 
but low expansion velocities, and PTF12gzk (Ben-Ami et al. 2012) that has extremely high expansion 
velocities but low velocity dispersion (and is therefore spectroscopically not a Ic-BL) requires usage
of velocity dispersion to define this class (rather than the easier to measure expansion velocity).  
With these exceptions noticed, the typical expansion velocity at peak for SNe Ic-BL is $19000$\,km\,s$^{-1}$
(Modjaz et al. 2016) and their mean $R$-band peak magnitude is M$_{\rm R}=-19.0$\,mag (Drout
et al. 2011).

\subsection{Other unusual Ib and Ic Supernovae}
\label{subsec:TypeIc:pec}

\subsubsection{Long rising events}
\label{subsubsec:TypeIc:pec:long}

Drout et al. (2011) demonstrated that SN Ib and SN Ic visible-light ($V$ and $R$ band) light curves
are relatively narrowly clustered around a mean curve. This finding is confirmed by Taddia et al. 
(2015) and Lyman et al.
(2016), in particular regarding the pre-peak, rising part of the light curve, 
and applies even for unusually luminous events such as SN 2010ay (Sanders et al. 2012).  
The large study by Prentice et al. (2016) shows a range of light curve shapes, but with rise times
from estimated explosion to peak in the range $10-22$\,d. Standard modelling relates the SN rise
time to its ejecta mass (Arnett 1982) and photospheric velocity. Since the velocities of normal 
SNe Ib and SNe Ic span a narrow range of values, the rise-time range implies a narrow range of 
low ejecta masses ($<10$\,M$_{\odot}$; e.g., Taddia et al. 2015).

A handful of events stand out as outliers, having broad light curves with much longer rise times; 
the most notable published example is SN 2011bm (Valenti et al. 2012) with a rise time of $35$\,d in the compilation of Prentice et al. (2016). Two recent works expand this small group with the inclusion 
of iPTF15dtg (Taddia et al. 2016b), which is very similar to SN 2011bm, and PTF11rka
(Pian et al. 2016, in preparation). The long rise times suggest much larger ejected masses,
that are consistent with the large masses of single Wolf-Rayet stars. The likely detection of a
shock-cooling signature in the early light curve of iPTF15dtg (and likely also SN 2011bm; Taddia 
et al. 2016b) provides strong evidence that this particular sub-class of SNe Ic arises from explosions
of massive stripped stars.  

Valenti et al. (2011) presented observations of SN 2009jf, a well-observed event with a remarkably long rise time ($\sim22$\,d in V band) compared, e.g., with the recent compilation of Prentice et al. 2016. They have classified it spectroscopically as a SN Ib, but comparison of its peak spectrum from Modjaz et al. (2014; Fig.~\ref{Fig_Ib_peculiar_peak}) shows it does not show
clear He I lines at peak and is very similar to the well-observed Type Ic SN 2007gr, as noted by Valenti et al. (2011). Based on this comparison, as well as on its location in Fig.~\ref{pEWfig}, it is definitely not a typical SN Ib; whether it is a long-rising SN Ic or some sort of hybrid/intermediate event is still unclear.   

\subsubsection{Late interactors}
\label{subsubsec:TypeIc:pec:int}

\begin{figure}[ht]
\sidecaption[t]
\includegraphics[width=120mm,trim=60 200 60 200 mm, clip=true]{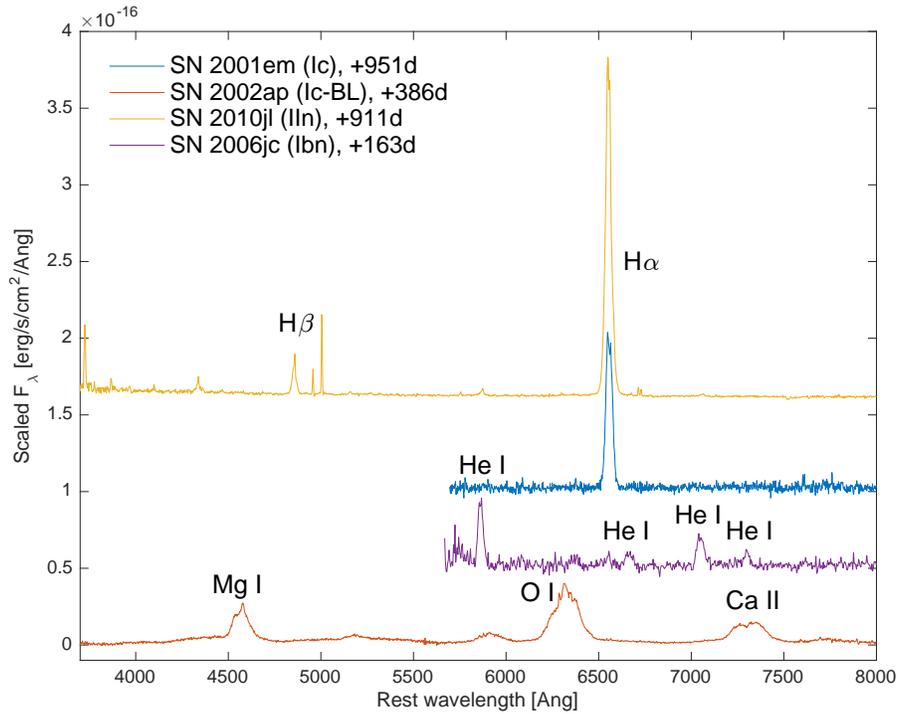}
%
\caption{A very late spectrum of the interacting Type Ic SN 2001em is dominated by 
hydrogen H$\alpha$ emission and shows striking similarity to late-time spectra of 
interacting SNe IIn (Soderberg et al. 2004), for example the 
late spectrum of the interacting Type IIn SN 2010jl (from Jencson et al. 2015). 
Broad nebular features of Ca, O and Mg, typical of late-time spectra of regular SNe Ic
(Fig.~\ref{Fig_I_neb}) are not seen, nor are narrow He I lines seen in late spectra of 
interacting Type Ibn SNe (a spectrum of SN 2006jc from Modjaz et al. 2014 is shown).}
\label{Fig_Ic_int}       
\end{figure}

Unlike Type Ibn ($\S$~\ref{subsubsec:TypeIb:pec:Ibn}) and Type IIn ($\S$~\ref{subsec:TypeII:IIn}) events, there is no evidence that regular SNe Ib and SNe Ic interact strongly with CSM. While some events show 
radio emission (e.g., Chevalier \& Soderberg 2010; Corsi et al. 2016) this emission 
presumably comes from
interaction of fast SN ejecta with low-mass material surrounding the explosion, and is not accompanied 
by signatures visible in optical spectra. A notable exception was 
SN 2001em, that showed bright
late-time radio emission (Granot \& Ramirez-Ruiz 2004), accompanied by prominent 
H$\alpha$ emission
in its late optical spectrum (Soderberg et al. 2004; Fig.~\ref{Fig_Ic_int}). While the SN was initially 
classified as a Type Ic event (Filippenko \& Chornock 2001) and this identification was adopted without question in the literature (e.g., Granot \& Ramirez-Ruiz 2004; Bietenholz \& Bartel 2005; Soderberg et al. 2006), a reanalysis of the original classification spectrum (shown
in Fig.~\ref{Fig_Ib_peculiar_peak}) kindly provided by Shivvers \& Filippenko, indicates SN 2001em
was actually a SN Ib. This likely indicates that the
SN ejecta caught up to, and were interacting with, dense, hydrogen-rich CSM. A similar case was
recently reported by Milisavljevic et al. (2015) and Margutti et al. (2016) for the Type Ib SN 2014C. Late-time radio and X-ray
emission from Type Ic-BL events (e.g., Corsi et al. 2014; 2016) may suggest this mode of strong
late-time interaction is also common to SNe Ic-BL.   

Ben-Ami et al. (2014) presented observations of SN 2010mb (PTF10iue), a unique SN Ic with
a remarkably extended, slowly-declining light curve (with a decline of $<2$\,mag in $R$-band over
$>500$\,d), inconsistent with being powered by radioactive $^{56}$Ni decay. The data show several signatures of strong interaction with CSM lacking H and probably also He, including the flat 
light curve shape, a blue spectral continuum, and transient narrow emission lines of 
neutral oxygen. The observations seem to require strong interaction with a large mass of
material ($\sim3$\,M$_{\odot}$) that surrounds
the progenitor. This event remains unique in the published literature both in terms of its light curve and
its spectroscopy. 
 
\subsubsection{Rapid decliners}
\label{subsubsec:TypeIc:pec:fast}

In large light curve studies of SNe Ib and SNe Ic (e.g., Drout et al. 2011, Bianco et al. 2014; Prentice et al. 2016) the spectroscopically normal Type Ic SN 1994I stands out as having a rapid decline from peak (approximately $1.5$\, mag in $R$-band within 15 days, or $\Delta$M$_{15,R}=1.5$\,mag). Three well-observed events show faster 
decline rates, and these have gained quite some attention: SN 2002bj (Poznanski et al. 2010), SN 2010X (Kasliwal et al. 2010) and SN 2005ek (Drout et al. 2013), all having similar values of 
$\Delta$M$_{15,R}=2.5-3$\,mag. The observations of SN 2002bj also constrain the rise time to be
very rapid, of order 7\,d. Spectroscopically, SN 2005ek and SN 2010X resemble spectroscopically normal SNe Ic, though they develop nebular features very rapidly (suggesting very low ejecta masses), while SN 2002bj may show helium features, though whether it resembles a normal SN Ib is unclear.
These events likely represent hydrogen-poor explosions with very low ejecta masses. If they are
all assumed to arise from the same progenitor, a massive star origin seems to be preferred, though
this is very much an open question still.   
 
Drout et al. (2014) presented a group of rapidly-declining events from the PS1 survey, with 
light curve properties similar to those of the rapidly-declining SNe Ib/c mentioned above, but
spectroscopically, observations are lacking to determine whether the PS1 events are SNe Ib/c
or rather rapidly-declining SNe IIn similar to PTF10uj (Ofek et al. 2010), or to the rapidly-evolving 
luminous class identified by
Arcavi et al. (2016). 

\section{Type II Supernovae}
\label{sec:TypeII}

This section reviews the classification of hydrogen-rich (Type II) SN events. For a detailed review of the
properties of the various subclasses defined here, see Chapter 3.2, ``H-rich Core-Collapse Supernovae''. 

\subsection{Regular Type II Supernovae}
\label{subsec:TypeII:reg}

\begin{figure}[ht]
\sidecaption[t]
\includegraphics[width=120mm,trim=50 200 50 200 mm, clip=true]{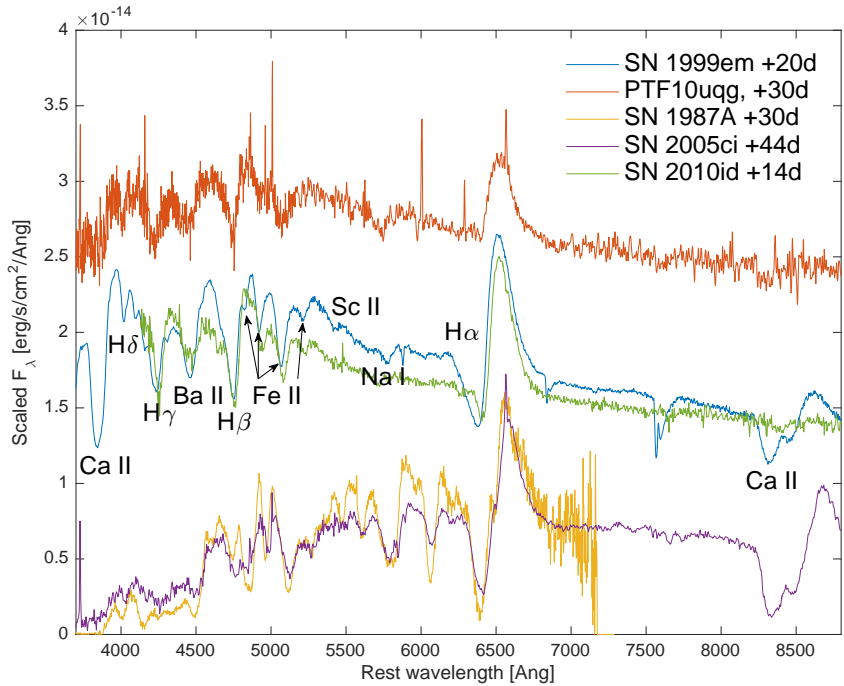}
\caption{Spectroscopically regular SNe in the photospheric phase. SN 1999em (blue; middle) is a well 
studied event (e.g., Leonard et al. 2002; spectrum shown is from Hamuy et al. 2001) with an extended
plateau (II-P). PTF10uqg (top) is the event with the steepest decline from peak 
($\Delta$m$_{15}=0.41$\,mag) among the regular events in the sample of Rubin et al. (2016), 
so a ``II-L'' like event; note the overall spectral similarity as well as the weaker absorption 
component of the H$\alpha$ line.
 At the bottom we show a spectrum of SN 1987A from Pun et al. (1995) at a similar age, as well as 
 a spectrum of another slowly-rising event, SN 2005ci from Taddia et al. (2016a), with more extended 
 red coverage; we consider events with such appearance to be spectroscopically regular SNe II.
 We also show a spectrum of the faint and slow Type II SN 2010id (PTF10vdl; olive; middle); note
 that even at 14d from estimated explosion this event has much lower expansion velocities. 
}
\label{Fig_II_regular_peak}       
\end{figure}

Spectroscopically-regular Type II SNe can be unambiguously defined using optical spectra taken
a few weeks after explosion, which are dominated by broad and strong lines of hydrogen (most
notably Balmer H$\alpha$) as seen in Fig.~\ref{Fig_II_regular_peak}. Spectroscopically-regular
SNe II show a range of light curve shapes, including objects with flat evolution in red light
(often called SNe II-P to denote their light curve Plateau), objects with relatively rapidly declining
light curves (referred to as SNe II-L) as well as object showing a prominent long rise, similar 
to SN 1987A. Chapter 3.2 describes these subclasses in detail and discusses the open issue of
whether they represent a continuum of properties or not. For the purpose of this section, we 
group all of these events together, as their spectroscopic appearance is quite similar. In Fig.~\ref{Fig_II_regular_peak} we identify the main spectroscopic features. A very detailed line identification
of a photospheric SN II spectrum is presented in Leonard et al. (2002; Fig. 10 and Table 4 in that paper). Rubin et al. (2016) find a mean peak absolute $R$-band magnitude of M$_{R}=-17.14$\,mag for spectroscopically normal events of all types. For the same sample of events, the expansion velocity
at peak as measured from the minimum of the H$\beta$ line is $9600$\,km\,s$^{-1}$.

\subsubsection{Spectroscopic Evolution}
\label{subsubsec:TypeII:reg:spec}

\begin{figure}[h]
\sidecaption[t]
\includegraphics[width=120mm,trim=50 70 50 60 mm, clip=true]{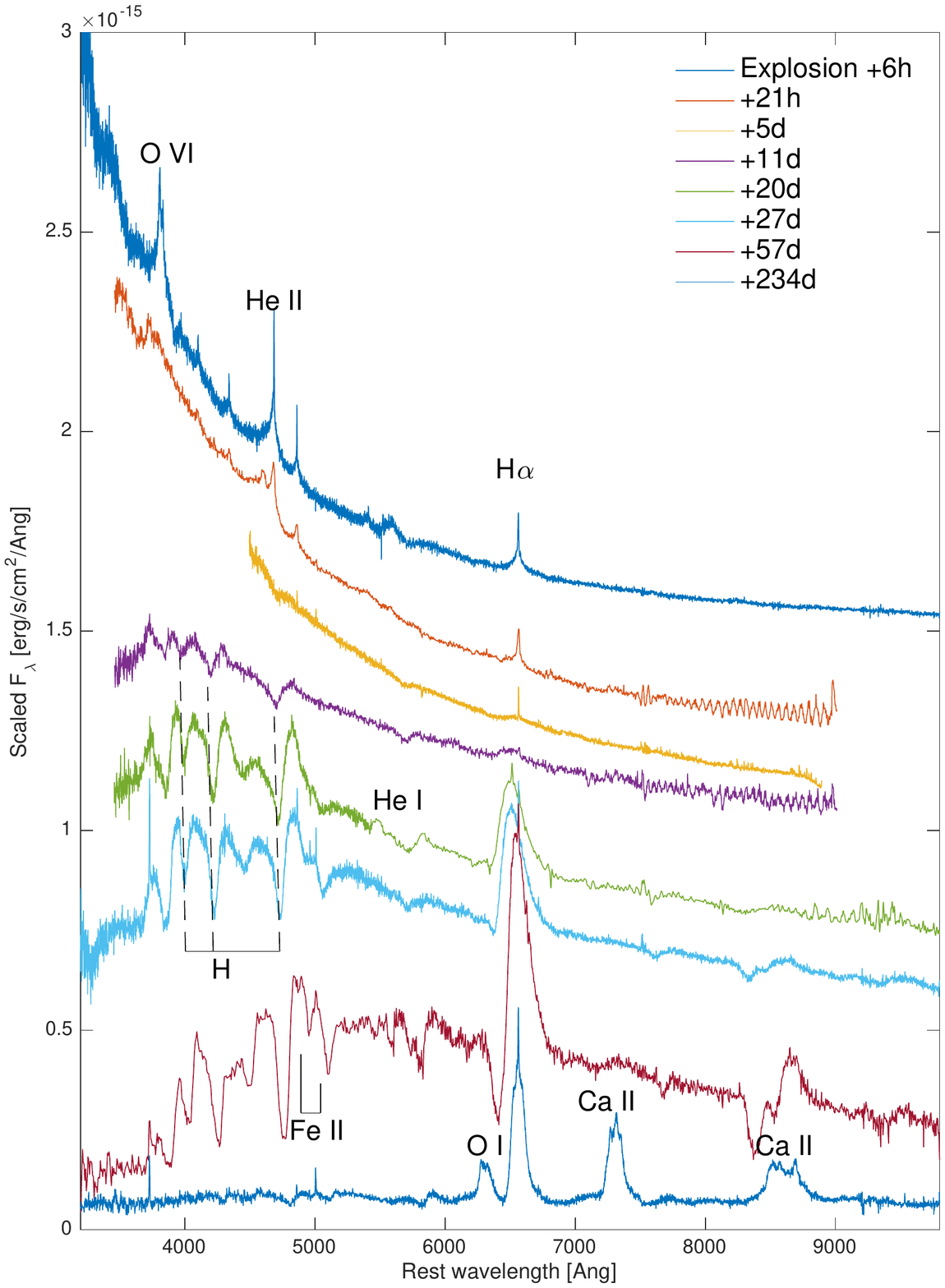}
\caption{Spectroscopic evolution of regular Type II SNe, exemplified by SN 2013fs (iPTF13dqy; Yaron et al. 2016). The spectra initially appear as very 
hot black-body-like curves (top). Spectra obtained within less than a day show high-excitation
emission lines (Gal-Yam et al. 2014; Yaron et al. 2016; top spectrum) while those obtained within
a few days often still show flash-ionized He II and Balmer emission lines (e.g., Khazov et al. 2015;
Terreran et al. 2016; Shivvers et al. 2015; Smith et al. 2015). 
These ``flash'' features typically disappear within
a few days of explosion, and the resulting blue and featureless spectra (third from top) are often seen
at discovery of SNe II. Within a week or two the spectrum transforms into the photospheric phase
with the emergence of strong, broad and blue-shifted He I  and Balmer lines (middle spectra). Within the next hundred days or so, the spectrum turns redder and strong metal lines appear (second from 
bottom spectrum). Eventually, the spectrum evolves into the nebular phase and is dominated by 
strong emission lines of H$\alpha$, Ca II and OI. 
}
\label{Fig_II_regular_specevol}       
\end{figure}

Fig.~\ref{Fig_II_regular_specevol} shows the evolution of the spectroscopically regular
SN 2013fs (iPTF13dqy; Yaron et al. 2016). This event has the most complete spectroscopic
coverage from hours to years after explosion. The figure illustrates the initial day-long flash-ionised
phase with prominent high-excitation emission lines, the transformation to a featureless blue continuum
commonly observed several days after explosion during the hot initial stage of the shock-cooling
phase, while hydrogen is likely fully ionised, the emergence of hydrogen lines at the
beginning of the photospheric phase and the subsequent evolution of additional metal lines, 
and the late-time nebular
phase. These phases are described in detail in the ``H-Rich Core-Collapse Supenrovae" chapter.  

\subsection{Slowly-Rising Type II}
\label{subsec:TypeII:87A}

The nearest SN in modern hostory, SN 1987A in the LMC, was a spectroscopically regular SN II
(Fig.~\ref{Fig_II_regular_peak}) with an exceedingly long rise time of $>80$\,days (Pun et al. 1995). This light curve peculiarity is 
rare, and during the decades since 1987 only a few additional examples were studied. Pastorello 
et al. (2012) and recently Taddia et al. (2016a) conducted studies of the small samples of such events
collected, while Pastorello et al. (2005), Kleiser et al. (2011), Taddia et al. (2012) and Arcavi et al. (2012) presented additional observations of a handful of individual objects. The compilation of 
Taddia et al. (2016a) suggests that these events have similar expansion velocities and peak magnitudes
as those of common SNe II (e.g., Rubin et al. 2016), except of course of the characteristic long
rise to peak. The progenitor of SN 1987A was observed (pre-explosion) to have been a blue supergiant star, and this
is likely true for all or most of the events in this group. 
 
\subsubsection{Faint and Slow Events}
\label{subsubsec:TypeII:reg:slow}

Pastorello et al. (2004) pointed out that a group of Type II SNe have substantially fainter
luminosities (typically peaking below M$_{R}=-15.5$\,mag) and lower expansion velocities 
(of order $50\%$ of the velocities measured for the prototypical SN 1999em, Fig.~\ref{Fig_II_regular_peak}; see Pastorello et al. 2009,
their Fig. 12), compared to the mean values given above (e.g., M$_{R}=-17.14$\,mag and
v$_{H\beta}=9600$\,km\,s$^{-1}$ from Rubin et al. 2016). SN 2005cs in the 
nearby M51 is a well-observed example (Pastorello et al. 2006; 2009), while SN 1999br
(Pastorello et al. 2004) and SN 2010id (PTF10vdl; Gal-Yam et al. 2011) are extreme events
(SN 2010id peaked at M$_{r}=-13.85$\,mag, Gal-Yam et al. 2011). In terms of their light-curve
shape, events of this group tend to have long and flat plateaus, and only occur within the 
fast rise slow decay (II-FS) subclass of Rubin \& Gal-Yam (2016). Since the progenitor masses
of these events have similar masses to those of regular SNe II (Smartt 2009), suggesting 
also similar ejecta masses, it is likely that the lower expansion velocities indicate these are 
sub-energetic (rather than higher mass) events. 

\subsection{Transitional Type IIb Events}
\label{subsec:TypeII:IIb}

Filippenko (1988) presented observations of SN 1987K that initially resembled Type II SNe, and 
then evolved to resemble a SN Ib, and proposed the term SN IIb to describe such transitional objects. 
The study of this class was boosted by the discovery of the nearby SN 1993J in M81 (e.g., 
Schmidt et al. 1993; Nomoto et al. 1993; Podsiadlowski et al. 1993; Swartz et al. 1993; Filippenko
et al. 1993). While SNe IIb are a relatively rare subclass ($5-10\%$ of SNe II; Arcavi et al. 2010;
Li et al. 2011; Ganot et al. 2016), several examples have been studied in great detail, including, 
in addition to SN 1993J mentioned above, also SN 2008ax (Pastorello et al. 2008c; Crockett
et al. 2008; Roming et al. 2009; Taubenberger et al. 2011) and the more recent 
SN 2011dh (Arcavi et al. 2011; 
Soderberg et al. 2012; Sahu et al. 2013; Horesh et al. 2013; Ergon et al. 2014; Marion et al. 2014;
Ergon et al. 2015) and SN 2013df (Van-Dyk et al. 2014; Morales-Garoffolo et al. 2014; Ben-Ami et al. 2015; Maeda et al. 2015). SN 2013cu (Gal-Yam et al. 2014) was discovered $<1$\,d after discovery
and has an extensive early spectroscopic and photometric coverage. Several SNe IIb have luminous massive 
progenitors according to pre-explosion {\it HST} data (e.g., Maund et al. 2004; 2011; Van Dyk et al. 2011; 2014).

This class of SNe is relatively well studied, with sample analysis papers focussed on radio properties
(Chevalier \& Soderberg et al. 2010), UV spectra (Ben-Ami et al. 2015), precursor activity (Strotjohann et al. 2015) and their optical properties (e.g., Arcavi et al. 2012; Liu et al. 2016 and Prentice et al. 2016). These recent works find mean peak magnitudes of M$_{R}=-17.94$\,mag (Prentice et al. 2016) and expansion velocities of $8400$\,km\,s$^{-1}$ (Liu et al. 2016) at peak. 

SN IIb light curves often show an initial early bump: famously for SN 1993J, and more recently for
e.g., SN 2011dh (Arcavi et al. 2011), SN 2011fu (Morales-Garoffolo et al. 2015) and SN 2013df
(Van Dyk et al. 2014). The light curve shape is usually similar to that of SNe Ib and Ic (e.g., Arcavi
et al. 2012), but a few events with more extended light curves have been identified 
(Rubin \& Gal-Yam 2016).

\begin{figure}[h]
\sidecaption[t]
\includegraphics[width=120mm,trim=50 200 50 200 mm, clip=true]{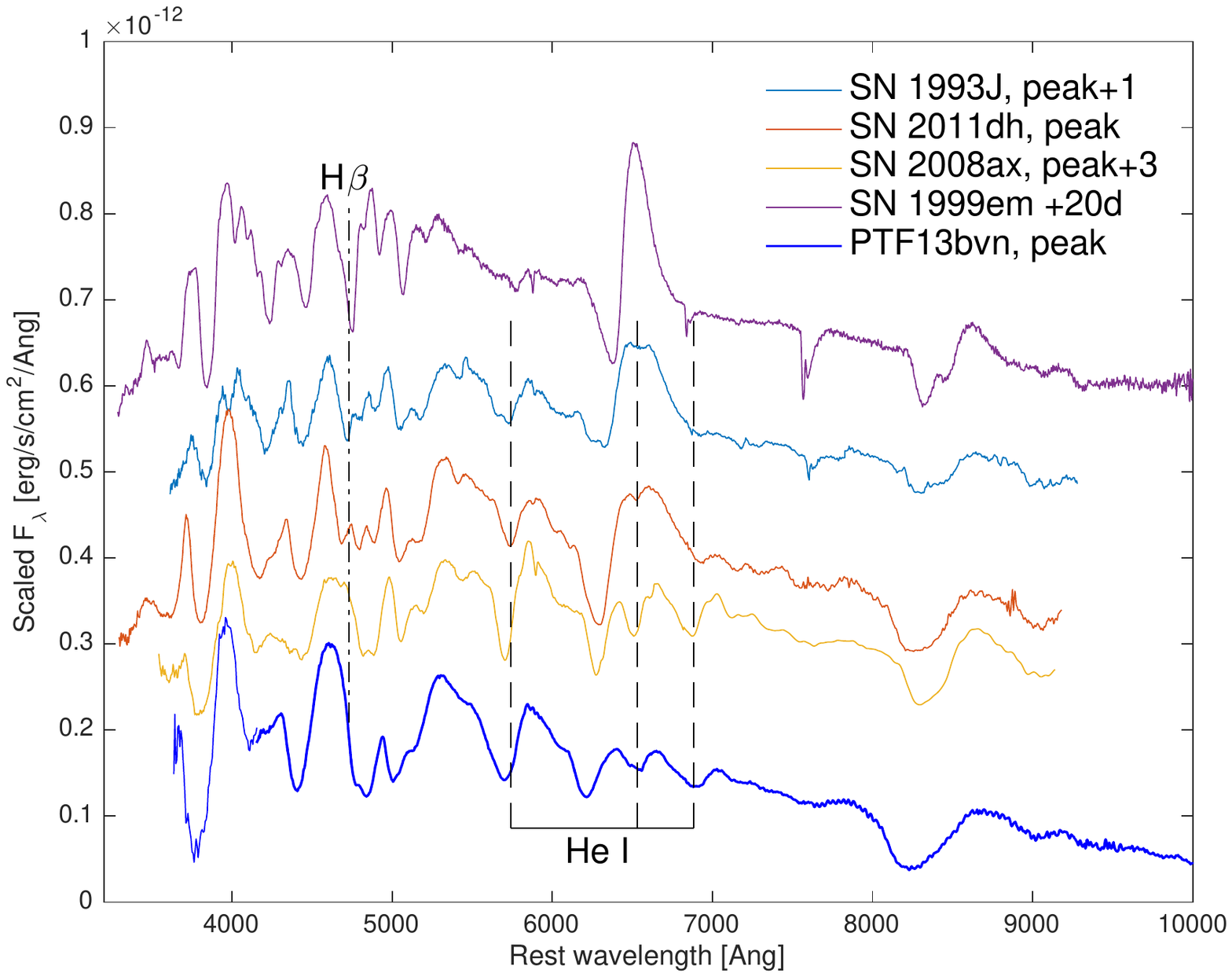}
\caption{SNe IIb as transition objects between regular SNe II (top, SN 1999em from Hamuy et al. 2011) and SNe Ib (bottom; PTF13bvn; see Fig.~\ref{Fig_Ib_peak}). Note the increasing prominence of the He I lines (dashed lines) in SNe IIb compared to regular SNe at the same age after explosion (top), as well as the gradual decrease in the strength of the hydrogen Balmer lines (dash-dot), with H$\beta$ becoming weak in the spectrum of SN 1993J (Barbon et al. 1995), barely detectable in the spectrum of  SN 2011dh (Ergon et al. 2014) and virtually absent in the spectrum of SN 2008ax (Taubenberger et al. 2011). The flat-top or ``notched'' profile of H$\alpha$ is a useful signature of SNe IIb, though one should take care to avoid contamination from telluric residuals or over-subtraction of narrow host galaxy Balmer lines.}
\label{Fig_IIb_peak}       
\end{figure}

As shown in Fig.~\ref{Fig_IIb_peak}, there is no clear spectroscopic separation between 
Type Ib and Type IIb SNe. Since extensive spectral sequences show that hydrogen in SNe IIb 
is typically 
limited only to the outer (faster) layers of the expanding ejecta (e.g. Ergon et al. 2014; their Fig. 14),
it may well be that SNe IIb and Ib represent a continuum of events with varying amounts
of residual hydrogen in their outer envelopes. Since it appears that all well-studies SNe Ib show
traces of hydrogen (e.g., Liu et al. 2016; Parrent et al. 2016), it may be better to reclassify all 
these events as members of a single class (IIb) and reserve the ``Ib'' class only for He-rich 
events that show no hydrogen at all. 

Several sub-classes of SNe IIb have been proposed over the years, including a radio-based 
distinction between events from compact vs. extended progenitors (Chevalier \& Soderberg 
2010) and a class of events with flat velocity evolution (Folatelli et al. 2014). 

\subsection{Interacting Type IIn Events}
\label{subsec:TypeII:IIn}

This subsection deals with the classification of hydrogen-rich SN events that show prominent
narrow emission lines (Type IIn). For a detailed review of the
properties of these events, see Chapter 3.6, ``Interacting Supernovae: Types IIn and Ibn''. 

The first examples of this class of events were presented by Filippenko (1989) and the term
``SN IIn'' introduced for the class by Schlegel (1990). Kiewe et al. (2012) presented a sample
of SNe IIn and reviewed the properties of all events published in the literature up to 2012. 
Stritzinger et al. (2012), Taddia et al. (2013) and Ofek et al. (2013; 2014a; 2016) has since presented
additional sample analysis papers. 

Chapter 3.6 below discusses in greater detail the interaction nature of SNe IIn, their 
origin in strong interaction with CSM surrounding the progenitors, and  
related subclasses (SLSN-II, Ibn and Ia-CSM). There is strong evidence that SNe IIn
arise from massive stars (e.g., Gal-Yam et al. 2007; Gal-Yam \& Leonard 2009; Smith et al. 2011).
Ofek et al. (2013; 2014b) showed that pre-explosion precursors appearing prior to the explosions
of SNe IIn are common. 

\begin{figure}[ht]
\sidecaption[t]
\includegraphics[width=120mm,trim=50 200 50 200 mm, clip=true]{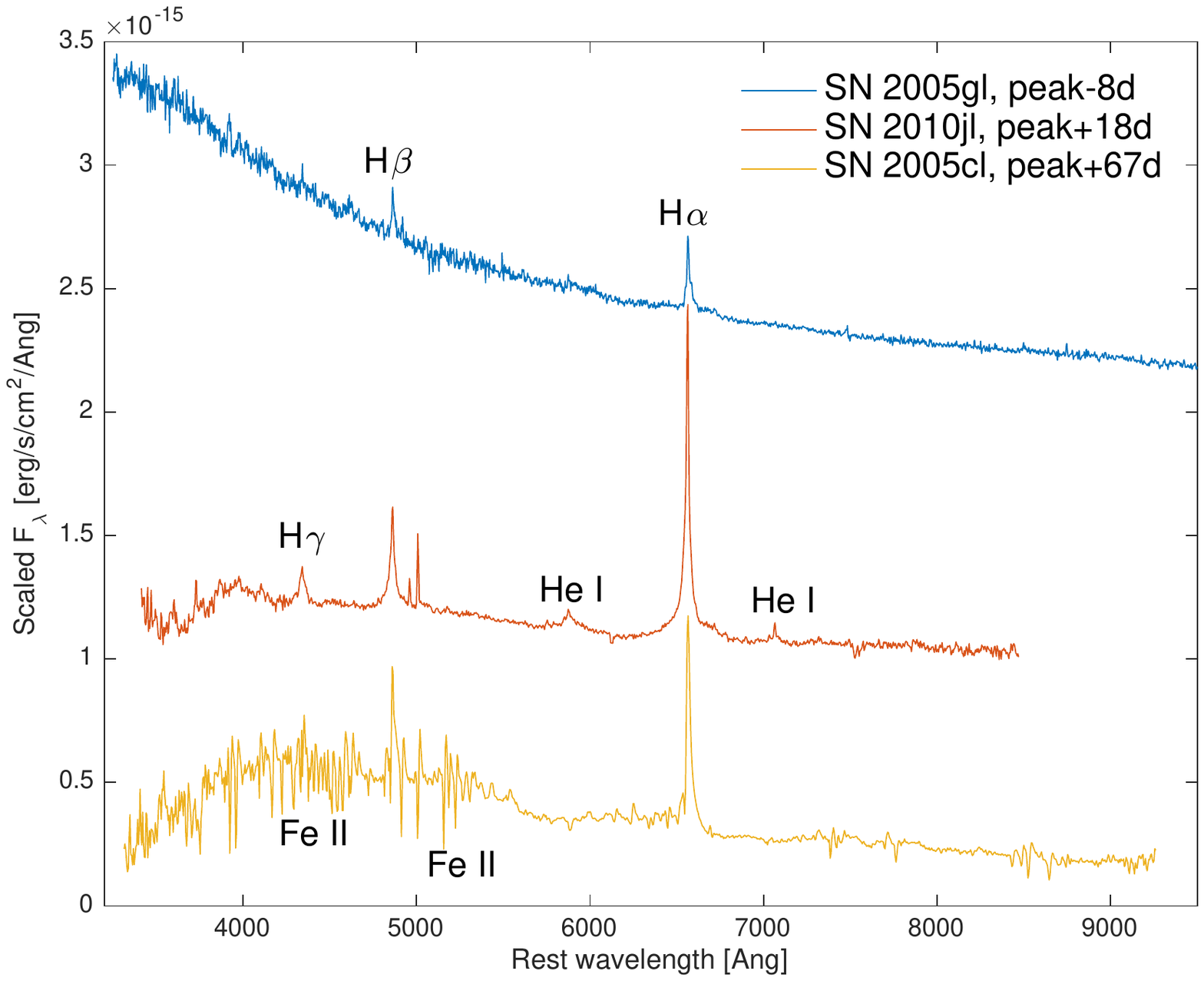}
\caption{Spectra of SNe IIn at pre-peak (SN 2005gl, Gal-Yam et al. 2005; top), post-peak (SN 2010jl, Zhang et al. 2012; middle) and moderately late phases (SN 2005cl, Kiewe et al. 2012; bottom). In view of the very diverse nature of this class of events these spectra only serve to illustrate some common features, including relatively smooth spectra with Balmer emission lines at early times, spectra dominated by strong Balmer and He I lines with extended wings (middle) and the devlopment of a blue pseudo-continuum that can sometimes be de-blended into numerous narrow lines of Fe II, e.g., bottom spectrum and Kiewe et al. (2012).}
\label{Fig_IIn_peak}       
\end{figure}

The photometric and spectroscopic properties of SNe IIn show a wide diversity. Fig.~\ref{Fig_IIn_peak} 
shows representative spectra. Since several classes of core-collapse SNe show transient 
flash-ionised emission lines during the first days after explosion (e.g., Gal-Yam et al. 2014; Khazov 
et al. 2016; see above) observations showing emission-line spectra that persist for $>10$\,d 
after explosion are needed to support a SN IIn classification. Kiewe et al. (2012) measure a
mean peak magnitude of M$_{R}=-18.77$\,mag for their sample of SNe IIn. Since the interaction
shock emission dominates the spectrum of SNe IIn at peak and obscures any photospheric 
features, it is not possible to measure the expansion velocities of these events. 
 
\section{SLSN}
\label{sec:SLSN}

The new class of superluminous supernovae (SLSNe) was defined (Gal-Yam 2012) 
during recent years, following the discovery of increasing numbers of extremely luminous
SNe, with peak magnitudes well in excess (typically $-21$\,mag absolute or above 
in visible light) of that of normal events (Table~\ref{tab:new-sys}). 
Gal-Yam (2012) provides a historical review of the emergence of this class.
Chapter 3.7 (''Superluminous Supernovae'') describes these events in more detail.

\begin{figure}[ht]
\sidecaption[t]
\includegraphics[width=120mm,trim=50 200 50 200 mm, clip=true]{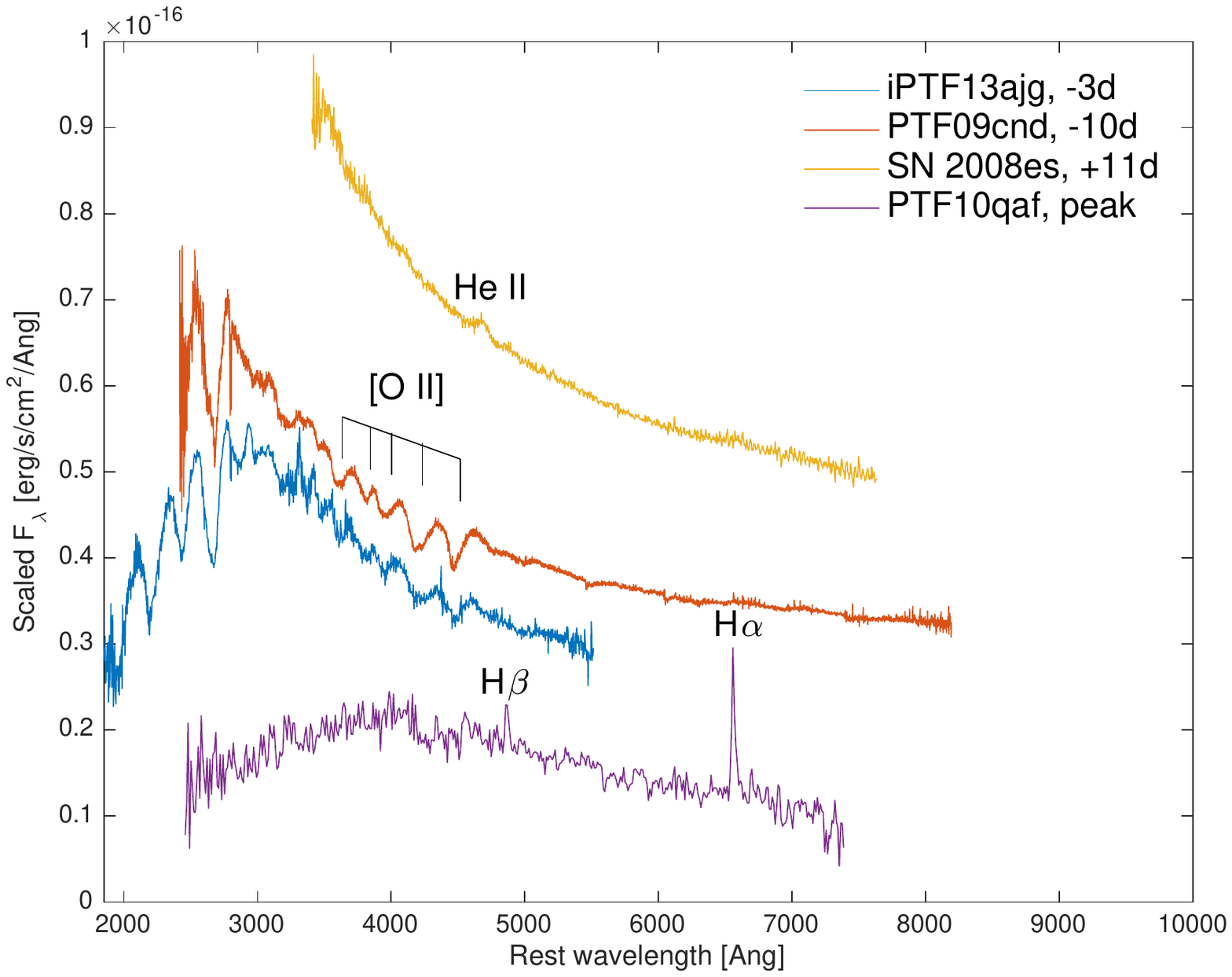}
\caption{Spectra of SLSNe around peak luminosity. SLSN-I typically show a ``comb'' of OII absorption lines in the optical (e.g., PTF09cnd, Quimby et al. 2011), as well as strong UV absorption features (prominent in the restframe UV spectra of iPTF13ajg from Vreeswijk et al. 2014) whose exact nature is still under debate. Common SLSNe-II show narrow Balmer emission lines, e.g., PTF10qaf (Gal-Yam 2012; Leloudas et al. in preparation) shown here. A minority of SLSNe-II exemplified by SN 2008es initially show blue, featureless spectra (with the exception of weak He II emission; e.g., Gezari et al. 2009), with hydrogen lines evolving only in later phases.} 
\label{Fig_SLSN_peak}       
\end{figure}

The classification of SLSNe broadly follows the nomenclature of regular SNe, with 
a basic division into Type II SLSNe (SLSNe-II) that show prominent hydrogen features,
and Type I events (SLSNe-I) that lack hydrogen (Fig.~\ref{Fig_SLSN_peak}; Gal-Yam 2012). 

The majority of SLSN-II 
events display strong hydrogen emission lines similar to those of the
less luminous SNe IIn (Fig.~\ref{Fig_SLSN_peak}; 
see Chapter 3.6, ``Interacting Supernovae: Types IIn and Ibn''
for more details). The nearest and best studied event of this class is SN 2006gy (e.g., Ofek et al. 2007;
Smith et al. 2007), with SN 2006tf (Smith et al. 2008), SN 2003ma (Rest et al. 2011) and SN 2008fz (Drake et al. 2010) serving as additional examples. It is unclear at this time whether 
SLSNe-II with narrow emission lines show spectral distinction from the most luminous SNe IIn,
see chapters 3.6 (``interacting supernovae'') and 3.7 (''Superluminous Supernovae'') for more
details. 

\begin{figure}[ht]
\sidecaption[t]
\includegraphics[width=120mm,trim=50 200 50 200 mm, clip=true]{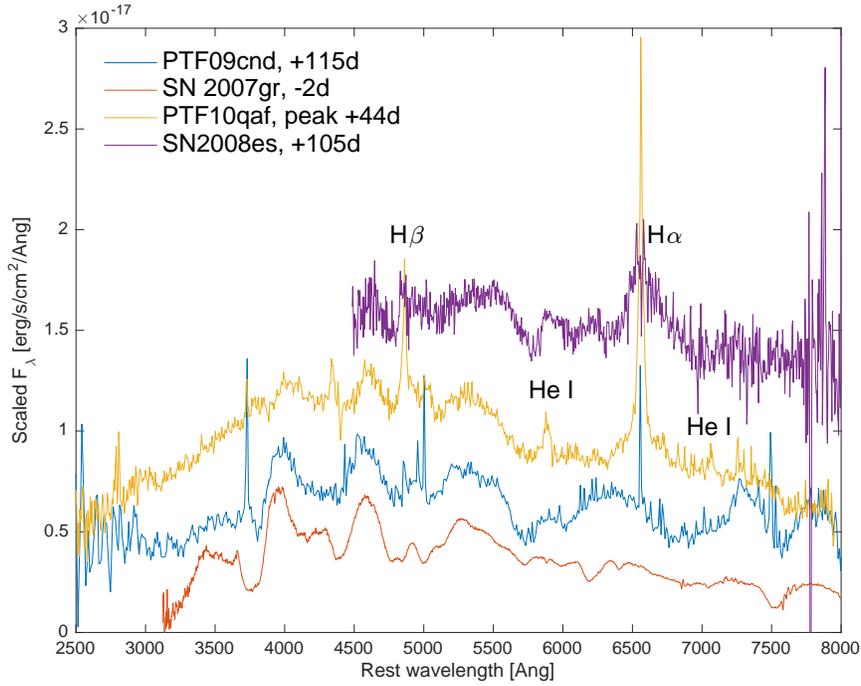}
\caption{Post-peak SLSN spectra. A few months after peak, SLSN-I spectra (e.g., PTF09cnd; Quimby et al. 2011) evolve to resemble those of SNe Ic (Pastorello et al. 2010; Quimby et al. 2011; a comparison spectrum of SN 2007gr from Valenti et al. 2008 is shown). SLSN-II such as SN 2008es (spectrum from Gezari et al. 2009; see also Miller et al. 2009) develop broad and prominent Balmer hydrogen lines. At least some SLSN-II with narrow lines (PTF10qaf from Gal-Yam 2012 shown here) present striking similarity post-peak to other SLSNe in their spectral shape, with superposed strong and sharp emission lines of the hydrogen Balmer series and He I.}
\label{Fig_SLSN_post}       
\end{figure}

A small number of SLSNe-II develop
broad hydrogen lines without narrow components (e.g., Fig.~\ref{Fig_SLSN_post}), 
the prototype of this class has been 
SN 2008es (Miller et al. 2009; Gezari et al. 2009). More recently Benetti et al. (2014) and Inserra 
et al. (2016b) presented additional examples and discussed their spectral similarity to 
SLSN-I. The late-time emergence of hydrogen lines in the spectra of SLSNe-I identified by
Yan et al. (2015) makes it difficult to determine whether these objects really constitute a separate
spectroscopic class. 

SLSNe-I have been defined as a spectroscopic class by Quimby et al. (2011), and 
are characterised by a unique initial phase showing a very blue continuum with a ``comb'' of absorption lines of OII in the optical (Quimby et al. 2011; Fig.~\ref{Fig_SLSN_peak}) and strong UV 
absorption features, in addition to their extreme luminosities. 
Later, these events evolve to resemble SNe Ic (Pastorello et al. 2010;
Quimby et al. 2011; Fig.~\ref{Fig_SLSN_post}).  
Interestingly, Fig.~\ref{Fig_SLSN_post} suggests that all types of SLSNe may share
spectroscopic similarities at late times.   

\begin{figure}[ht]
\sidecaption[t]
\includegraphics[width=120mm,trim=50 200 50 200 mm, clip=true]{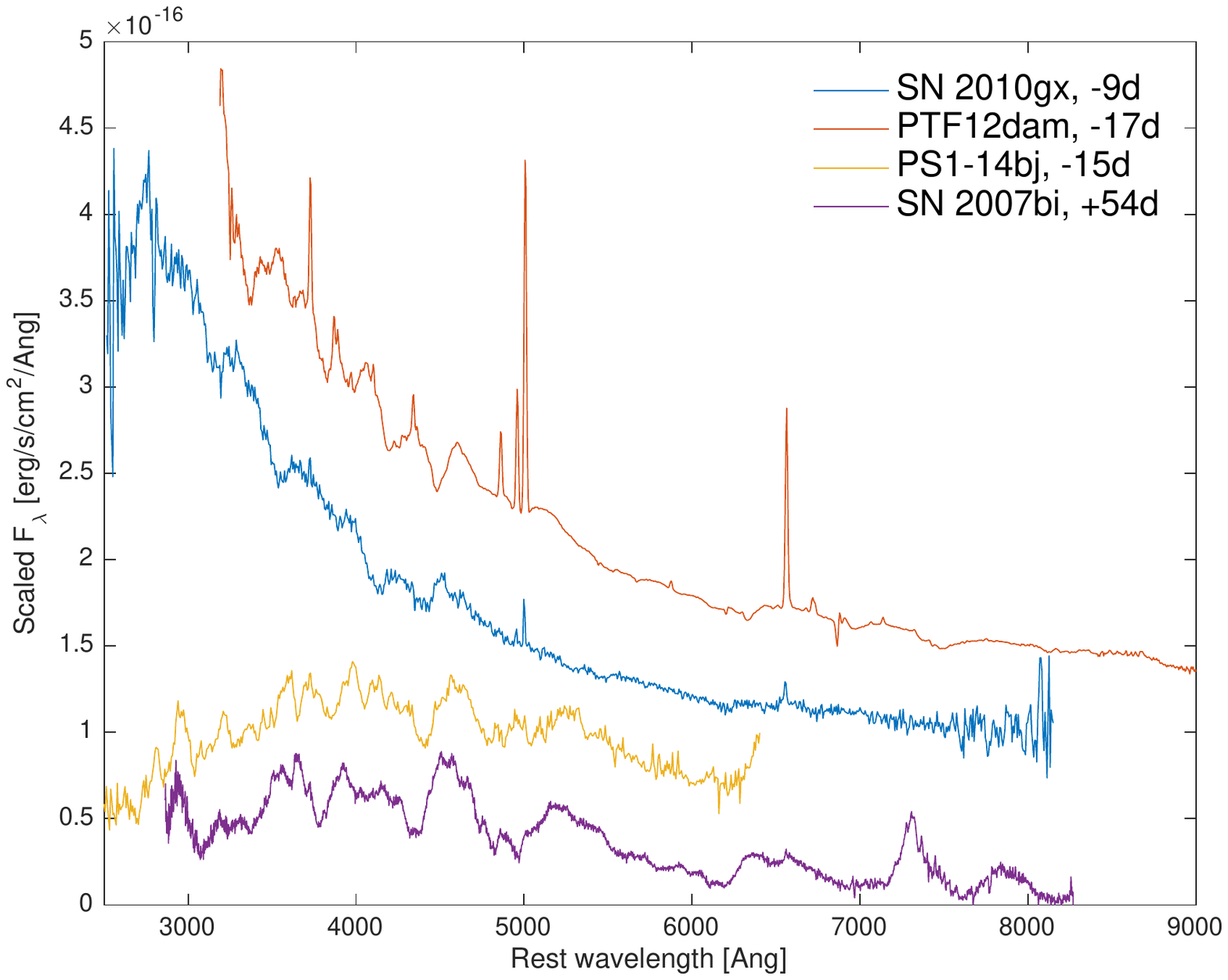}
\caption{Spectra of SLSN-I before peak. Some events show the ``standard'' hot spectrum prior
to peak, with a blue continuum and the OII ``comb'', these include both rapidly declining
(e.g., SN 2010gx, Pastorello et al. 2010; spectrum shown is from Quimby et al. 2011), and
slowly declining (PTF12dam; spectrum from Nicholl et al. 2013). Some events show red and
evolved spectra even prior to peak, e.g., PS1-14bj (Lunnan et al. 2016), the examples known
so far are only slowly-declining events similar to SN 2007bi (Gal-Yam et al. 2009).}
\label{Fig_SLSN_I_sub}       
\end{figure}

In terms of their light curve evolution, SLSNe-I are better studied than SLSNe-II. In addition
to their extreme peak luminosities, they often show initial ``bumps'' in their light curves
(Leloudas et al. 2011; Nicholl et al. 2015a; Smith et al. 2016; Nicholl \& Smartt 2016); these
may be related to the explosion shock breakout cooling emission. Some events also show
prominent bumps at later times (e.g., Nicholl et al. 2016). Gal-Yam (2012) suggested that the
class of SLSNe-I could be separated into subclasses of slowly (Type R) and rapidly (Type I) 
declining events, but additional observations accumulated since makes this distinction 
unclear. Spectroscopic evidence (Fig.~\ref{Fig_SLSN_I_sub}) may suggest that at least some 
slowly- and rapidly-declining SLSNe-I also show spectroscopic distinction, but more
pre-peak spectroscopic observations of SLSNe-I are required to better test this. 

Since SLSNe are detected throughout a broad redshift range and via surveys with diverse 
observing strategies and typical depths, it is not trivial to define their mean properties. De Cia 
et al. (in preparation) use the complete sample of PTF-discovered SLSNe-I (e.g., Perley et al. 2016)
to measure a mean restframe $g$-band magnitude of M$_{g}=-21.1$\,mag 
that translates to a mean $r$-band peak magnitude of M$_{r}=-21.5\pm0.8$\,mag, where
the dispersion includes both dispersion in peak magnitude and peak $g-r$ color. The
mean expansion velocities of SLSNe have not yet been well characterised. No similar
analysis has been carried out yet for SLSNe-II.

While no direct evidence regarding the nature of SLSN progenitors exist, numerous lines of 
evidence (for example, slow rise and decline, which require extreme ejecta masses; e.g., Gal-Yam
et al. 2009; Nicholl et al. 2015b) as well as their host-galaxy properties (e.g., Neill et al. 2011;
Lunnan et al. 2014; 2015; Leloudas et al. 2015; Perley et al. 2016) suggest the progenitors 
are massive stars.   

\section{A New Supernova Classification Scheme}
\label{sec:new}

The existing SN classification scheme is based mostly on peak spectroscopic features for the main
classes, and involves also other properties for sub-classes. The system evolved over several decades
and new classes and sub-classes were added as our knowledge expanded, with nomenclature
often taking some years to stabilise (in the sense of most people using the same term for the
same group of objects). With the maturation of the field, now may be the time to undertake a review
of this issue and propose a modified classification scheme. This is attempted here. 

A few shortcomings of the existing system one can think of is that class names are not systematically
defined according to a set of underlying principles, some of the class names are quite opaque to non-experts and form a barrier for general astrophysicists in following the SN literature, and that some class names mix observational properties with physical interpretation that may be debated (for example, ``Ia-CSM'').
 
The new system presented here is inspired by the system used to classify stars. The main principles followed are: 

1. Class names clearly separate assumptions about the physical origin of events 
(that are often a subject of debate) and 
observational properties (that should be universally agreed upon, to within measurement errors).

2. The system tries to retain maximal backward compatibility, in particular with regards to 
the class of SNe Ia that has become a ``brand'' among the general scientific community.

3. The system is based on observable quantities that are (and will remain) most readily measured,
namely peak magnitude and peak optical spectra.

4. Optional tags are introduced, that allow for short names for standard objects, and more 
informative and long names for unusual objects.

5. The system can easily accommodate continuous (rather than discrete) class designations,
to handle events located along a continuum of some defining property.

6. The system is well suited to implementation within large data bases.\\

The new SN class names have the following structure: 

\begin{equation}
{\rm SN}{\underbar A}{\underbar A}~{\rm X.Y~iX.Y~mX.Y~vX.Y~rX.Y~dX.Y}
\end{equation}
   
All SN class names have a required part composed of a physical origin indicator {\underbar A}{\underbar A} with values \underline{AA}= ``Ia'' for explosions of WD progenitors, and \underline{AA}=``CC'' for 
massive star explosions. This is followed by a spectroscopically defined classifier of the composition
of the ejecta, with ``0'' standing for strong hydrogen features (no or little stripping of the progenitor),
``1'' stands for strong He features with no H features (all hydrogen stripped), and ``2'' indicates no
He features (all H and He stripped). This system easily accommodates transitional classes such
as the group currently named SNe IIb, which would have fractional values between 0 and 1. Since 
standard SNe Ia never show hydrogen or helium features, the composition indicator can also be
dropped for this class, defaulting to SN Ia (rather than SN Ia2) for standard objects.     

This required part is followed by optional tags. These include ``i'' for interaction, followed by the 
composition of the material with which the SN ejecta interact (i0 = interaction with hydrogen, i1 = 
interaction with helium, and so on). The ``m'' tag indicates the peak magnitude offset from the 
typical value for a particular class, and the ``v'' tag indicates fractional shift in velocity with 
respect to the mean peak expansion velocity. We also introduce two tags to note the temporal 
scales of objects. the tag ``r'' notes the rise time
to peak in days, and the tag ``d'' notes the decline rate in magnitudes per 15 days in $R$-band  
($\Delta$M$_{R,15}$).

Table~\ref{tab:new-sys} provides a full translation
between currently used classes defined above and the new system.
Peak magnitudes have been homogenised to the $R$ band.

\begin{table}
\caption{A New Classification System for Supernovae}
\label{tab:new-sys}       
\begin{tabular}{p{2.3cm}p{1.7cm}p{1.2cm}p{2.0cm}p{3.5cm}}
\hline\noalign{\smallskip}
Old Class & Physical     & Peak          & Peak Expansion & New Class \\
                 & Origin$^a$ & $R$-band   & Velocity              &  \\
                 &                   & [mag]         & [km\,s$^{-1}$]     & \\
\noalign{\smallskip}\svhline\noalign{\smallskip}
SN Ia & WD & -18.67 & 11000 & SNIa(2)\\
\noalign{\smallskip}
SN Ia (91T-like) & WD? & -19.15 & 11000 & SNIa2\,m-0.5\\
\noalign{\smallskip}
SN Ia (91bg-like) & WD & -17.55 & 11000 & SNIa2\,m+1.1\\
\noalign{\smallskip}
SN Ia (Super-C) & WD? & $\sim$-19.5 & 8000 & SNIa2\,m-1\,v0.7\\
\noalign{\smallskip}
SN Ia (02cx-like) & WD? & -17 & $<8000$ & SNIa2\,m+1.6\,v0.7\\
\noalign{\smallskip}
SN Ia-CSM & WD? & -20.2 &  & SNIa2i0\,m-1.5\\
\noalign{\smallskip}
SN Ib & Massive star  & -17.9 & 10000 & SNCC0.9 \\
\noalign{\smallskip}
SN 2005bf & Massive star?  & -18 & 7000 & SNCC1\,v0.7 \\
\noalign{\smallskip}
Ca-rich Ib & ?  & -16 & 10000 & SN??1\,m+2 \\
\noalign{\smallskip}
SN Ibn & Massive star  & -19 &  & SNCC2\,i1 \\
\noalign{\smallskip}
SN Ibn/IIn$^b$ & Massive star  &   &  & SNCC2\,i0.5 \\
\noalign{\smallskip}
SN Ic  & Massive star  & -18.3 & 10000 & SNCC2 \\
\noalign{\smallskip}
SN Ic-BL & Massive star & -19 & 19000 & SNCC2\,m-0.7\,v1.9\\
\noalign{\smallskip}
Long-rising Ic & Massive star &   &   & SNCC2\,r35\\
\noalign{\smallskip}
Late interacting Ic$^c$ & Massive star? &   &   & SNCC2\,i0\\
\noalign{\smallskip}
SN 2010mb$^d$ & Massive star? &   &   & SNCC2\,d0\\
\noalign{\smallskip}
Rapid decliners$^e$ & Massive star? &   &   & SNCC2\,d2.5\\
\noalign{\smallskip}
Regular SNe II  & Massive star & 17.1 & 10000 & SNCC0\\
\noalign{\smallskip}
SNe II-P  & Massive star &  &  & SNCC0\,d0\\
\noalign{\smallskip}
SNe II-L  & Massive star &  &  & SNCC0\,d0.3\\
\noalign{\smallskip}
Long-rising II$^f$  & Massive star &  &  & SNCC0\,r80\\
\noalign{\smallskip}
Faint II$^g$  & Massive star & -15 & 5000 & SNCC0\,m+1.5\,v0.5\\
\noalign{\smallskip}
SN IIb  & Massive star & -18 & 8500 & SNCC0.5\\
\noalign{\smallskip}
SN IIn  & Massive star & -18.8 & & SNCC0\,i0\\
\noalign{\smallskip}
SLSN-I  & Massive star? & -21.5 &  & SNCC2\,m-3.2\\
\noalign{\smallskip}
SLSN-II  & Massive star? &  &  & SNCC0\,m-2\\
\noalign{\smallskip}
SLSN-IIn & Massive star? &  &  & SNCC0\,i0\,m-2\\
\noalign{\smallskip}\hline\noalign{\smallskip}
\end{tabular}
\\
$^a$ Values with ``?'' attempt to reflect the consensus in the literature, even though direct evidence is not available. \\
$^b$ Events similar to SN 2005la (Pastorello et al. 2008b)\\
$^c$ Events similar to SN 2001em and SN 2014C ($\S$~3.5.2)\\
$^d$ Ben-Ami et al. 2012, see $\S$~3.5.2\\
$^e$ Events similar to SNe 2002bj, 2010X and 2005ek ($\S$~3.5.3)\\
$^f$ SN 1987A-like\\
$^g$ $\S$~4.2.1\\

\end{table}

\begin{acknowledgement}
The author acknowledges support from the Kimmel Award. I thank A. Rubin,
M. Sullivan, A. V. Filippenko and I. Shivvers for contributing data, analysis and 
advice. This work benefited enormously from that WISeREP spectroscopic 
data base (Yaron \& Gal-Yam 2012) that would not have been made possible without
the vision, creativity and skill of O. Yaron, the hard work of I. Manulis, and the numerous
member of the supernova research community that made their data publicly available, for which 
I deeply thank them.
 
\end{acknowledgement}

%
%
%

\begin{thebibliography}{99.}%


\bibitem{phys-journal} Arnett, W.~D. et al.\ 1982, ApJ, 253, 785 
\bibitem{phys-journal} Arcavi, I. et al.\ 2010, ApJ, 721, 777 
\bibitem{phys-journal} Arcavi, I. et al.\ 2011, ApJ, 742, L18 
\bibitem{phys-journal} Arcavi, I. et al.\ 2012, ApJ, 756, L30 
\bibitem{phys-journal} Arcavi, I. et al.\ 2016, ApJ, 819, 35 
\bibitem{phys-journal} Ashall, C. et al.\ 2016, MNRAS, 460, 3529 


\bibitem{phys-journal} Barbon, R. et al.\ 1979, A\&A, 72, 287 
\bibitem{phys-journal} Barbon, R. et al.\ 1995, A\&AS, 110, 513 
\bibitem{phys-journal} Ben-Ami, S. et al.\ 2012, ApJ, 760, L33 
\bibitem{phys-journal} Ben-Ami, S. et al.\ 2014, ApJ, 785, 37 
\bibitem{phys-journal} Ben-Ami, S. et al.\ 2015, ApJ, 803, 40 
\bibitem{phys-journal} Benetti, S. et al.\ 2006, ApJ, 653, L129 
\bibitem{phys-journal} Benetti, S. et al.\ 2011, MNRAS, 411, 2726 
\bibitem{phys-journal} Benetti, S. et al.\ 2014, MNRAS, 441, 289 
\bibitem{phys-journal} Bianco, F.~B. et al.\ 2014, ApJS, 213, 19 
\bibitem{phys-journal} Bietenholz, M.~F. \& Bartel, N.\ 2005, ApJ, 625, L99 
\bibitem{phys-journal} Blondin, S. et al.\ 2012, AJ, 143, 126 
\bibitem{phys-journal} Bloom, J.~S. et al.\ 2012, ApJ, 744, L17 


\bibitem{phys-journal} Cao, Y. et al.\ 2013, ApJ, 775, L7 
\bibitem{phys-journal} Cao, Y. et al.\ 2015, Nature, 521, 328
\bibitem{phys-journal} Cao, Y. et al.\ 2016, ApJ, submitted, arXiv:1606.05655
\bibitem{phys-journal} Chevalier, R.~A. \& Soderberg, A.~M.\ 2010, ApJ, 711, L40
\bibitem{phys-journal} Childress, M.~J. et al.\ 2013, ApJ, 770, 29 
\bibitem{phys-journal} Chornock, R. et al.\ 2011, ApJ, 739, 41
\bibitem{phys-journal} Chugai, N. N.. \& Danziger, I. J.\ 1994, MNRAS, 268, 173 
\bibitem{phys-journal} Clocchiatti, A. et al.\ 2006, ApJ, 642, 1 
\bibitem{phys-journal} Corsi, A. et al.\ 2012, ApJ, 747, L5 
\bibitem{phys-journal} Corsi, A. et al.\ 2014, ApJ, 782, 42 
\bibitem{phys-journal} Corsi, A. et al.\ 2016, ApJ, submitted, arXiv:1512.01303 
\bibitem{phys-journal} Crockett, R.~M. et al.\ 2008, MNRAS, 391, L5


\bibitem{phys-journal} Dessart, L. et al.\ 2012, MNRAS, 424, 2139 
\bibitem{phys-journal} Dilday, B. et al.\ 2012, Science, 337, 942 
\bibitem{phys-journal} Drake, A.~J. et al.\ 2010, ApJ, 718, L127 
\bibitem{phys-journal} Drout, M.~R. et al.\ 2011, ApJ, 741, 97 
\bibitem{phys-journal} Drout, M.~R. et al.\ 2013, ApJ, 774, 58 


\bibitem{phys-journal} Elias, J.~H. et al.\ 1985, ApJ, 296, 379 
\bibitem{phys-journal} Ergon, M. et al.\ 2014, A\&A, 562, 17 
\bibitem{phys-journal} Ergon, M. et al.\ 2015, A\&A, 580, 142 


\bibitem{phys-journal} Filippenko, A.~V.\ 1988, AJ, 96, 1941 
\bibitem{phys-journal} Filippenko, A.~V.\ 1989, AJ, 97, 726 
\bibitem{phys-journal} Filippenko, A.~V. et al.\ 1992a, ApJ, 384, L37 
\bibitem{phys-journal} Filippenko, A.~V. et al.\ 1992b, AJ, 104, 1543 
\bibitem{phys-journal} Filippenko, A.~V. et al.\ 1993, ApJ, 415, L103 
\bibitem{phys-journal} Filippenko, A.~V.\ 1997, ARA\&A, 35, 309 
\bibitem{phys-journal} Filippenko, A.~V. \& Chornock, R.,\ 2001, IAUC, 7737 
\bibitem{phys-journal} Folatelli, G. et al.\ 2006, ApJ, 641, 1039 
\bibitem{phys-journal} Folatelli, G. et al.\ 2014, ApJ, 792, 7 
\bibitem{phys-journal} Foley, R.~J. et al.\ 2003, PASP, 115, 1220 
\bibitem{phys-journal} Foley, R.~J. et al.\ 2007, ApJ, 657, L105 
\bibitem{phys-journal} Foley, R.~J. et al.\ 2013, ApJ, 767, 57 
\bibitem{phys-journal} Foley, R.~J. et al.\ 2016, MNRAS, 461, 433


\bibitem{phys-journal} Galama, T.~J. et al.\ 1998, Nature, 395, 670 
\bibitem{phys-journal} Gal-Yam, A. et al.\ 2002, MNRAS, 332, L73 
\bibitem{phys-journal} Gal-Yam, A. et al.\ 2007, ApJ, 656, 372 
\bibitem{phys-journal} Gal-Yam, A. \& Leonard, D.~C.\ 2009, Nature, 458, 865 
\bibitem{phys-journal} Gal-Yam, A., et al.\ 2009, Nature, 462, 624 
\bibitem{phys-journal} Gal-Yam, A. et al.\ 2011, ApJ, 736, 159 
\bibitem{phys-journal} Gal-Yam, A.\ 2012, Science, 337, 927 
\bibitem{phys-journal} Gal-Yam, A., et al.\ 2014, Nature, 509, 471 
\bibitem{phys-journal} Ganot, A., et al.\ 2016, ApJ, 820, 57 
\bibitem{phys-journal} Ganeshalingam, M. et al.\ 2012, ApJ, 751, 142 
\bibitem{phys-journal} Gezari, S. et al.\ 2009, ApJ, 690, 1313 
\bibitem{phys-journal} Gorbikov, E. et al.\ 2014, MNRAS, 443, 671 
\bibitem{phys-journal} Granot, J. \& Ramirez-Ruiz, E.\ 2004, ApJ, 609, L9 


\bibitem{phys-journal} Hachinger, S. et al.\ 2012, MNRAS, 422, 70 
\bibitem{phys-journal} Hamuy, M. et al.\ 2001, ApJ, 558, 615 
\bibitem{phys-journal} Hamuy, M. et al.\ 2003, Nature, 424, 651 
\bibitem{phys-journal} Hicken, M. et al.\ 2007, ApJ, 669, L17 
\bibitem{phys-journal} Horesh, A. et al.\ 2013, MNRAS, 436, 1258 
\bibitem{phys-journal} Hosseinzadeh, G. et al.\ 2016, ApJ, submitted 
\bibitem{phys-journal} Howell, D.~A. et al.\ 2001, ApJ, 554, L193 
\bibitem{phys-journal} Howell, D.~A. et al.\ 2006, Nature, 443, 308 
\bibitem{phys-journal} Harkness, R.~P. et al.\ 1987, ApJ, 317, 355 


\bibitem{phys-journal} Inserra, C. et al.\ 2016a, MNRAS, 459, 2721
\bibitem{phys-journal} Inserra, C. et al.\ 2016b, MNRAS, in press, arXiv:1604.08207 


\bibitem{phys-journal} Jencson, J.~E. et al.\ 2016, MNRAS, 456, 2622 


\bibitem{phys-journal} Kasliwal, M.~M. et al.\ 2008, ApJ, 683, L29 
\bibitem{phys-journal} Kasliwal, M.~M. et al.\ 2010, ApJ, 723, L98 
\bibitem{phys-journal} Kasliwal, M.~M. et al.\ 2012, ApJ, 755, 161 
\bibitem{phys-journal} Khazov, D. et al.\ 2016, ApJ, 818, 3 
\bibitem{phys-journal} Kiewe, M. et al.\ 2012, ApJ, 744, 10 
\bibitem{phys-journal} Kleiser, I.~K. et al.\ 2011, MNRAS, 415, 372 


\bibitem{phys-journal} Leloudas, G. et al.\ 2012, A\&A, 541, 129
\bibitem{phys-journal} Leloudas, G. et al.\ 2015, MNRAS, 449, 917
\bibitem{phys-journal} Leloudas, G. et al.\ 2015, A\&A, 574, 61
\bibitem{phys-journal} Leonard, D.~C. et al.\ 2012, PASP, 114, 35
\bibitem{phys-journal} Li, W. et al.\ 2001, ApJ, 546, 734
\bibitem{phys-journal} Li, W. et al.\ 2001, PASP, 113, 1178
\bibitem{phys-journal} Li, W. et al.\ 2003, PASP, 115, 453
\bibitem{phys-journal} Li, W. et al.\ 2011, MNRAS, 412, 1441
\bibitem{phys-journal} Liu, Y. et al.\ 2016, ApJ, in press, arXiv:1510.08049 
\bibitem{phys-journal} Lunnan, R. et al.\ 2014, ApJ, 787, 138
\bibitem{phys-journal} Lunnan, R. et al.\ 2015, ApJ, 804, 90
\bibitem{phys-journal} Lunnan, R. et al.\ 2016, ApJ, submitted, arXiv:1605.05235
\bibitem{phys-journal} Lyman, J.~D. et al.\ 2016, MNRAS, 457, 328 


\bibitem{phys-journal} Maeda, K. et al.\ 2007, ApJ, 666, 1069 
\bibitem{phys-journal} Maeda, K. et al.\ 2015, ApJ, 807, 35 
\bibitem{phys-journal} Maguire, K. et al.\ 2011, MNRAS, 418, 747 
\bibitem{phys-journal} Maguire, K. et al.\ 2014, MNRAS, 444, 3258 
\bibitem{phys-journal} Margutti, R. et al.\ 2016, ApJ, submitted, arXiv:1601.06806 
\bibitem{phys-journal} Marion, G.~H. et al.\ 2014, ApJ, 781, 69 
\bibitem{phys-journal} Matheson, T. et al.\ 2001, AJ, 121, 1648 
\bibitem{phys-journal} Maund, J.~R. et al.\ 2004, Nature, 427, 129 
\bibitem{phys-journal} Maund, J.~R. et al.\ 2011, ApJ, 739, L37 
\bibitem{phys-journal} Maund, J.~R. et al.\ 2007, MNRAS, 381, 201 
\bibitem{phys-journal} Mazzali, P.~A. et al.\ 1995, A\&A, 297, 509 
\bibitem{phys-journal} Mazzali, P.~A. et al.\ 1997, MNRAS, 284, 151 
\bibitem{phys-journal} Mazzali, P.~A. et al.\ 2002, ApJ, 572, L61 
\bibitem{phys-journal} Mazzali, P.~A. et al.\ 2014, MNRAS, 439, 1959 
\bibitem{phys-journal} Miller, A.~A. et al.\ 2009, ApJ, 690, 1303 
\bibitem{phys-journal} Milisavljevic, D.\ 2015, ApJ, 815, 120 
\bibitem{phys-journal} Minkowski, R.\ 1941, PASP, 53, 224 
\bibitem{phys-journal} Modjaz, M. et al.\ 2009, ApJ, 702, 226 
\bibitem{phys-journal} Modjaz, M. et al.\ 2014, AJ, 147, 99 
\bibitem{phys-journal} Modjaz, M. et al.\ 2016, ApJ, in press, arXiv:1509.07124
\bibitem{phys-journal} Morales-Garoffolo, A. et al.\ 2014, MNRAS, 445, 1647
\bibitem{phys-journal} Morales-Garoffolo, A. et al.\ 2015, MNRAS, 454, 95


\bibitem{phys-journal} Nakar, E.\ 2015, ApJ, 807, 172 
\bibitem{phys-journal} Neill, J.~D.\ 2011, ApJ, 727, 15 
\bibitem{phys-journal} Nicholl, M. et al.\ 2013, Nature, 502, 346 
\bibitem{phys-journal} Nicholl, M. et al.\ 2015a, ApJ, 807, L18 
\bibitem{phys-journal} Nicholl, M. et al.\ 2015b, MNRAS, 452, 3869 
\bibitem{phys-journal} Nicholl, M. et al.\ 2016, ApJ, 826, 39 
\bibitem{phys-journal} Nicholl, M., \& Smartt, S.~J.\ 2016, MNRAS, 457, L79 
\bibitem{phys-journal} Nomoto, K. et al.\ 1993, Nature, 364, 507
\bibitem{phys-journal} Nugent, P.~E.et al.\ 2011, Nature, 480, 344 


\bibitem{phys-journal} Ofek, E.~O.et al.\ 2017, ApJ, 659, L13 
\bibitem{phys-journal} Ofek, E.~O.et al.\ 2010, ApJ, 724, 1396 
\bibitem{phys-journal} Ofek, E.~O.et al.\ 2013, ApJ, 763, 42 
\bibitem{phys-journal} Ofek, E.~O.et al.\ 2013, Nature, 494, 65 
\bibitem{phys-journal} Ofek, E.~O.et al.\ 2014, ApJ, 788, 154 
\bibitem{phys-journal} Ofek, E.~O.et al.\ 2014, ApJ, 789, 104 
\bibitem{phys-journal} Ofek, E.~O.et al.\ 2016, ApJ, submitted 


\bibitem{phys-journal} Pan, Y.-C. et al.\ 2015, MNRAS, 452, 4307
\bibitem{phys-journal} Parrent, J.~T. et al.\ 2016, ApJ, 820, 75
\bibitem{phys-journal} Pastorello, A. et al.\ 2004, MNRAS, 347, 74 
\bibitem{phys-journal} Pastorello, A. et al.\ 2005, MNRAS, 360, 950 
\bibitem{phys-journal} Pastorello, A. et al.\ 2006, MNRAS, 370, 1752 
\bibitem{phys-journal} Pastorello, A. et al.\ 2007, Nature, 447, 829 
\bibitem{phys-journal} Pastorello, A. et al.\ 2008a, MNRAS, 389, 113
\bibitem{phys-journal} Pastorello, A. et al.\ 2008b, MNRAS, 389, 131 
\bibitem{phys-journal} Pastorello, A. et al.\ 2008c, MNRAS, 389, 955 
\bibitem{phys-journal} Pastorello, A. et al.\ 2009, MNRAS, 394, 2266 
\bibitem{phys-journal} Pastorello, A. et al.\ 2010, ApJ, 724, L16 
\bibitem{phys-journal} Pastorello, A. et al.\ 2012, A\&A, 537, A141 
\bibitem{phys-journal} Pastorello, A. et al.\ 2016, MNRAS, 456, 853 
\bibitem{phys-journal} Perets, H.~B. et al.\ 2010, Nature, 465, 322 
\bibitem{phys-journal} Perley, D.~A. et al.\ 2016, ApJ, in press, arXiv:1601.08207 
\bibitem{phys-journal} Podsiadlowski, Ph. et al.\ 1993, Nature, 364, 509 
\bibitem{phys-journal} Poznanski, D. et al.\ 2010, Science, 327, 58
\bibitem{phys-journal} Prentice, S.~J. et al.\ 2016, MNRAS, 458, 2973
\bibitem{phys-journal} Pun, J. et al.\ 1995, ApJS, 99, 223 


\bibitem{phys-journal} Quimby, R.~M. et al.\ 2011, Nature, 474, 487 


\bibitem{phys-journal} Rajala, A.~M. et al.\ 2005, PASP, 117, 132 
\bibitem{phys-journal} Rest, A. et al.\ 2011, ApJ, 729, 88 
\bibitem{phys-journal} Roming, P.~W.~A. et al.\ 2009, ApJ, 704, L118
\bibitem{phys-journal} Rubin, A. et al.\ 2016, ApJ, 820, 33
\bibitem{phys-journal} Rubin, A. \& Gal-Yam A.\ 2016, ApJ, in press, arXiv:1602.01446
\bibitem{phys-journal} Ruiz-Lapuente, P. et al.\ 1992, ApJ, 387, L33
\bibitem{phys-journal} Ruiz-Lapuente, P. et al.\ 1993, Nature, 365, 728


\bibitem{phys-journal} Sahu, D.~K.. et al.\ 2013, MNRAS, 433, 2 
\bibitem{phys-journal} Sanders, N.~E.. et al.\ 2012, ApJ, 756, 184 
\bibitem{phys-journal} Scalzo, R.~A. et al.\ 2010, ApJ, 713, 1073 
\bibitem{phys-journal} Schlegel, E.~M.\ 1990, MNRAS, 244, 269 
\bibitem{phys-journal} Schmidt, B.~P. et al.\ 1993, Nature, 364, 600 
\bibitem{phys-journal} Shivvers, I.\ 2015, ApJ, 806, 213 
\bibitem{phys-journal} Silverman, J.~M. et al.\ 2011, MNRAS, 425, 1789 
\bibitem{phys-journal} Silverman, J.~M. et al.\ 2012, MNRAS, 425, 1819 
\bibitem{phys-journal} Silverman, J.~M. et al.\ 2013a, ApJS, 207, 3 
\bibitem{phys-journal} Silverman, J.~M. et al.\ 2013b, ApJ, 772, 125 
\bibitem{phys-journal} Silverman, J.~M. et al.\ 2013c, MNRAS, 436, 1225 
\bibitem{phys-journal} Smartt, S.~J.\ 2009, ARA\&A, 47, 63 
\bibitem{phys-journal} Smith, M., et al.\ 2016, ApJ, 818, L8 
\bibitem{phys-journal} Smith, N., et al.\ 2007, ApJ, 666, 1116 
\bibitem{phys-journal} Smith, N., et al.\ 2008, ApJ, 686, 467 
\bibitem{phys-journal} Smith, N., et al.\ 2011, ApJ, 732, 63 
\bibitem{phys-journal} Smith, N.\ 2014, ARA\&A, 139, 1451 
\bibitem{phys-journal} Smith, N., et al.\ 2015, MNRAS, 449, 1876 
\bibitem{phys-journal} Soderberg, A.~M. et al.\ 2004, GCN 2586 
\bibitem{phys-journal} Soderberg, A.~M. et al.\ 2006, ApJ, 638, 930 
\bibitem{phys-journal} Soderberg, A.~M. et al.\ 2008, Nature, 453, 469 
\bibitem{phys-journal} Soderberg, A.~M. et al.\ 2012, ApJ, 752, 78 
\bibitem{phys-journal} Stritzinger, M. et al.\ 2009, ApJ, 696, 713 
\bibitem{phys-journal} Stritzinger, M. et al.\ 2012, ApJ, 756, 173 
\bibitem{phys-journal} Sullivan, M. et al.\ 2011, ApJ, 732, 118 
\bibitem{phys-journal} Swartz, D.~A. et al.\ 1993, Nature, 365, 232 


\bibitem{phys-journal} Taddia, F. et al.\ 2012, A\&A, 537, A140 
\bibitem{phys-journal} Taddia, F. et al.\ 2013, A\&A, 555, 10 
\bibitem{phys-journal} Taddia, F. et al.\ 2015, A\&A, 574, 60 
\bibitem{phys-journal} Taddia, F. et al.\ 2016a, A\&A, 588, A5 
\bibitem{phys-journal} Taddia, F. et al.\ 2016b, A\&A, submitted, arXiv:1605.02491 
\bibitem{phys-journal} Tanaka, M. et al.\ 2010, ApJ, 714, 1209 
\bibitem{phys-journal} Taubenberger, S. et al.\ 2006, MNRAS, 371, 1459 
\bibitem{phys-journal} Taubenberger, S. et al.\ 2009, MNRAS, 397, 677 
\bibitem{phys-journal} Taubenberger, S. et al.\ 2011, MNRAS, 412, 2735 
\bibitem{phys-journal} Taubenberger, S. et al.\ 2011, MNRAS, 413, 2140 
\bibitem{phys-journal} Taubenberger, S. et al.\ 2013, ApJ, 775, L43 
\bibitem{phys-journal} Terreran, G. et al.\ 2016, MNRAS, submitted, arXiv:1605.06116
\bibitem{phys-journal} Tominaga N. et al.\ 2005, ApJ, 633, L97 


\bibitem{phys-journal} Valenti, S. et al.\ 2008, ApJL, 673, L155 
\bibitem{phys-journal} Valenti, S. et al.\ 2011, MNRAS, 416, 3138 
\bibitem{phys-journal} Valenti, S. et al.\ 2012, ApJ, 749, L28 
\bibitem{phys-journal} Van Dyk, S.~D. et al.\ 2011, ApJ, 741, L28 
\bibitem{phys-journal} Van Dyk, S.~D. et al.\ 2014, AJ, 147, 37 


\bibitem{phys-journal} Waxman, E. et al.\ 2007, ApJ, 667, 351 
\bibitem{phys-journal} Wheeler, J.~C. \& Levreault, R.\ 1985, ApJ, 294, L17 
\bibitem{phys-journal} Wheeler, J.~C. \& Harkness, R.~P.\ 1990, RPPh, 53, 1467 
\bibitem{phys-journal} White, C.~J. et al.\ 2015, ApJ, 799, 52 
\bibitem{phys-journal} Woosley, S.~E. \& Bloom, J.~S.\ 2006, ARA\&A, 44, 507 


\bibitem{phys-journal} Yan, L. et al.\ 2015, ApJ, 814, 108 
\bibitem{phys-journal} Yaron, O.~\& Gal-Yam, A.\ 2012, PASP, 124, 668 
\bibitem{phys-journal} Yaron, O. et al.\ 2016, Nature Physics, submitted 


\bibitem{phys-journal} Zhang, T. et al.\ 2012, AJ, 144, 131 

%
%
%
%
%
%
%
\bigskip

%
%
%
%
%
%
\bigskip
%

\end{thebibliography}
%

\end{document}